\documentclass[prb,twocolumn,superscriptaddress,showpacs]{revtex4-1}

\usepackage{amsmath,amssymb}
\usepackage[dvipdfmx]{graphicx}
\usepackage{bm}
\usepackage{bbm}
\usepackage[hypertex,colorlinks,bookmarks=true,citecolor=blue,linkcolor=black,urlcolor=magenta]{hyperref}
\usepackage{color}
\usepackage{srcltx} 
\usepackage{times,txfonts}

\allowdisplaybreaks

\begin{document}

\title{Laughlin's argument for the quantized thermal Hall effect
}
\date{\today}
\author{Ryota Nakai}
\affiliation{WPI-Advanced Institute for Materials Research (WPI-AIMR), Tohoku University,
Sendai 980-8577, Japan}
\email{rnakai@wpi-aimr.tohoku.ac.jp}
\author{Shinsei Ryu}
\affiliation{Department of Physics, University of Illinois, 
1110 West Green St, Urbana IL 61801}
\author{Kentaro Nomura}
\affiliation{Institute for Materials Research,
Tohoku University, Sendai 980-8577, Japan}

\begin{abstract}
We extend Laughlin's magnetic-flux-threading argument to the quantized thermal Hall effect.
A proper analogue of Laughlin's adiabatic magnetic-flux threading process
for the case of the thermal Hall effect is given in terms of an external gravitational field.
From the perspective of the edge theories of quantum Hall systems,   
the quantized thermal Hall effect is closely tied to
the breakdown of large diffeomorphism invariance, 
that is, a global gravitational anomaly.
In addition, we also give an argument from the bulk perspective
in which a free energy, decomposed into its Fourier modes,
is adiabatically transferred under an adiabatic process involving external gravitational perturbations.
\end{abstract}


\maketitle

\tableofcontents

\section{Introduction}
\label{sec:introduction}

The thermal Hall conductivity is quantized 
in gapped $(2+1)$-dimensional topological phases \cite{prange87,hasan10,qi11} 
of charged and charge-neutral excitation systems.
Integer and fractional quantum Hall systems \cite{kane97} and chiral p-wave topological superconductors \cite{read00} are examples of such systems.
More precisely, the thermal Hall conductivity 
in these systems is given by 
\begin{align}
 \kappa_{H}
 =
 c\frac{\pi k_{\text{B}}^2T}{6\hbar},
\end{align}
where $c$ is the chiral central charge of the gapless boundary modes.
Hence, $\kappa_H$ is quantized in units of 
$\pi k_{\text{B}}^2T/6\hbar$.
For example, 
an integer quantum Hall system with the bulk Chern number $\nu$ of the filled electronic energy bands has $\nu$ complex-fermionic boundary modes
with $c=\nu$, 
and a topological superconductor with the Chern number $\nu$ of the Bogoliubov quasiparticles has $\nu$ Majorana boundary modes 
with $c=\nu/2$.

The quantized thermal Hall effect in two-dimensional topological insulators and topological superconductors (superfluids) has been discussed 
both from bulk and boundary points of view. 
From the perspective of chiral gapless boundary theories,
the thermal Hall effect has been studied in terms of  
the chiral Luttinger liquid \cite{kane97},
the conformal field theory \cite{cappelli02,bradlyn15-2},
the gravitational Chern-Simons theory \cite{stone12}, 
and the equilibrium partition function \cite{nakai16}.
On the other hand, the thermal Hall effect in the quantum Hall bulk is much controversial.
Various studies using 
the Kubo formula \cite{smrcka77,cooper97,qin11},
the non-equilibrium Green's function \cite{shitade14},
and the St\v{r}eda formula \cite{nomura12}
have concluded that the bulk fermionic states show the quantized thermal Hall effect.
However,
from the point of view of equilibrium thermal field theories, 
the thermal Hall current in the bulk is
exponentially small when the temperature is much smaller than the bulk energy gap \cite{manes13}.
Also,  an induced gravitational field theory derived from a fully gapped fermionic system in a thermal equilibrium cannot describe the quantized thermal Hall effect \cite{bradlyn15}.
These results may imply that, 
while for chiral edge theories one can develop 
an argument for the quantized thermal Hall effect, parallel to the quantum Hall effect,
the bulk picture of the quantized thermal Hall effect may be distinct from that for the quantum Hall effect.

In this paper, we extend 
the gauge invariance/noninvariance argument presented by Laughlin \cite{laughlin81}
to the thermal Hall effect in quantum Hall systems. 
Laughlin's argument provides a fundamental and robust theory of adiabatic responses in gapped topological phases.
We will make an attempt to follow as closely as possible the original Laughlin's argument,
by making one-to-one correspondence 
between electromagnetism and gravity (or more precisely, not full Einstein 
gravity but gravitoelectromagnetism).
We will discuss the adiabatic responses of the chiral boundary
fermion modes and the bulk quantum Hall states 
against the gravitational counterpart of the magnetic-flux threading.

From the edge-theoretical point of view, 
we elucidate the role of quantum anomalies connecting the boundary theories and Laughlin's argument. 
In particular, we will make use of the {\it global} gravitational anomaly of 
the boundary theories, as opposed to the {\it perturbative} gravitational anomaly.
While the perturbative gravitational anomaly correctly accounts for the non-conservation of the energy-momentum of the chiral edge theories, and hence the necessity of having 
the bulk system, 
it is not entirely obvious how one could relate the non-conservation of the energy-momentum to the thermal transport.
As we will discuss, the connection to the thermal transport is more transparent if we base our discussion
on the global gravitational anomaly.
It should however be noted that
the global gravitational anomaly, 
i.e., the anomalous phase of the partition function, 
has an ambiguity $2\pi\times \mbox{integer}$.
One may then worry that 
the global gravitational anomaly may not have an ability to 
fix the thermal transport coefficient entirely.
Nevertheless,
this ambiguity can be lifted by requiring 
consistency with the perturbative gravitational anomaly.

As for the bulk point of view,   
our thermal extension of Laughlin's argument 
reveals a picture quite analogous 
to the quantized charge Hall current that flows adiabatically 
through the bulk, i.e., the creeping of Landau orbitals as one threads a magnetic flux
adiabatically. 
In particular, 
to explain the thermal Hall effect, 
it seems that it is possible to avoid the use of non-equilibrium frameworks,
and confine our discussion entirely within the thermal effective 
field theory,
as in other anomaly-related transport phenomena.

This paper is organized as follows.
In Sec.\ \ref{Laughlin's original argument for the quantum Hall effect},
we start by reviewing the Laughlin's original argument of flux-threading. 
In Sec.~\ref{sec:boundary_qhe}, 
Laughlin's argument is recast into the language of 
the chiral boundary theories. 
In particular, 
we distinguish 
two types of quantum anomalies, 
perturbative and global U(1) gauge anomalies.
While at the level of the quantized charge transport, 
both anomalies lead to the same conclusion
(the quantized Hall effect), 
the distinction between the perturbative and global anomalies 
is an important prerequisite for the later application.  
In Sec.~\ref{sec:boundary_the}, we first show that the flux threading in the gravitational case is described by a modular transformation of the base manifold. Then the thermal Hall effect is explained by a global gravitational anomaly regarding the modular invariance of the boundary theory.
In Sec.~\ref{sec:bulk}, the thermal Hall effect is quantitatively explained 
from the bulk point of view. 
Finally in Sec.~\ref{sec:conclusion}, we summarize our results.

\section{Bulk argument for the quantum Hall effect
(Laughlin's original argument)
} 
\label{Laughlin's original argument for the quantum Hall effect}

\begin{figure}[t]
 \centering
 \includegraphics[width=48mm]{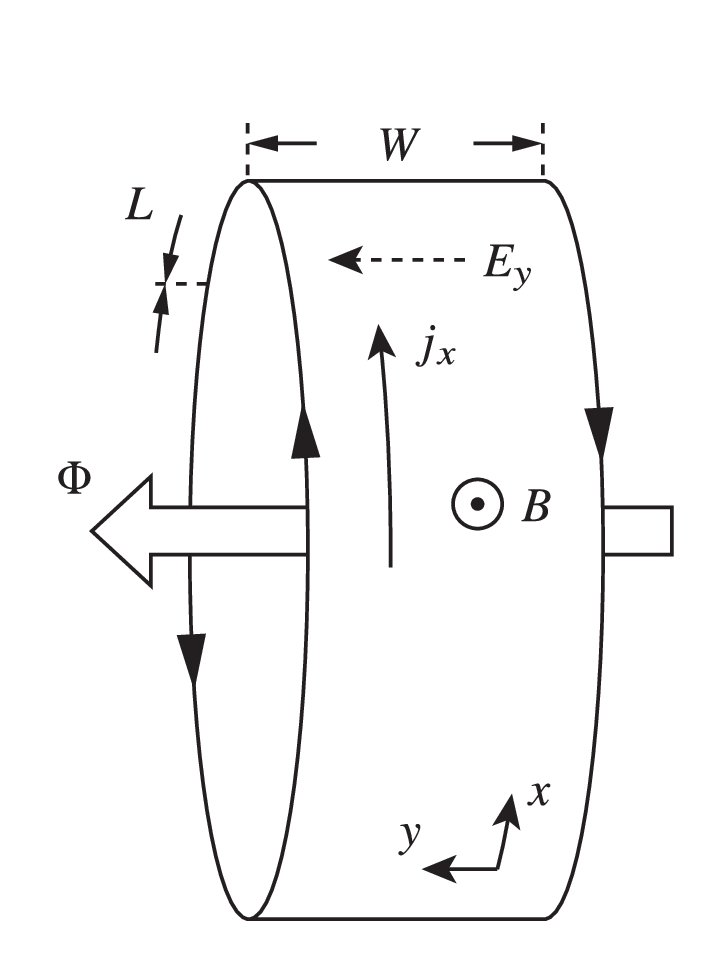}
 \caption{The cylindrical geometry for Laughlin's argument. 
 Electrons are confined on the cylindrical surface 
 in the presence of a magnetic field $B$
 applied perpendicular to the surface. 
 A magnetic flux $\Phi$ is threaded through 
 the hole of the cylinder, and an electric field $E_y$ is applied.\label{fig:laughlin_geometry}}
\end{figure}

Let us start by reviewing  
some notations and fundamentals of 
the quantum Hall effect
by following the Laughlin's original argument. 
Laughlin's argument explains 
the quantized Hall effect from the {\it bulk} point of view.
Consider an electronic system confined on the cylindrical surface (Fig.~\ref{fig:laughlin_geometry})
of the $x$-$y$ plane.
A magnetic field $B>0$ is applied in the out-of-plane ($z$) direction. 
Consider a two-dimensional electron gas ($e<0$) described by a quadratic single-particle Hamiltonian
\begin{align}
 \mathcal{H}
 =
 \frac{1}{2m}
 \left(
 -i\hbar
 \bm{\partial}
 -
 e\bm{A}
 \right)^2,
 \label{eq:2deg}
\end{align}
with the Landau gauge vector potential $\bm{A}=(-By,0)$, which is consistent with the periodic boundary in the $x$ direction.
The electronic system has translational invariance in the $x$ direction, and thus eigenstates are labeled by the wave number $k_x$.
The Hamiltonian (\ref{eq:2deg}) 
has the discrete energy spectrum consisting of the Landau levels,
\begin{align}
 \epsilon_{N}
 =
 \hbar\omega_c
 \left(
 N
 +
 \frac{1}{2}
 \right),
 \label{eq:landaulevel_energy}
\end{align}
where $N$ is a non-negative integer labeling the Landau levels, 
and $\omega_c=|e|B/m$ is the cyclotron frequency.
When the Fermi level lies in the energy gap between the Landau levels $\nu$ and $\nu+1$, 
that is, 
the eigenstates up to the Landau level $\nu$ are occupied, 
the electrons below the Fermi level carry a quantized Hall 
conductivity as $\sigma_{\text{H}}=\nu e^2/2\pi\hbar$.
The eigenstate wave functions are given by 
\begin{align}
 \phi_{N,k_x}(x,y)
 \propto
 e^{ik_x x}
 e^{-(y-y_0)^2/2l^2}
 H_N(y-y_0),
 \label{eq:qh_wavefunction}
\end{align}
where $l=(\hbar/|e|B)^{1/2}$ is the magnetic length 
and $H_N(y)$ is the Hermite polynomial of degree $N$. 
The wave functions (\ref{eq:qh_wavefunction}) are localized 
in the $y$ direction about a point $y_0$, 
and extended in the $x$ direction.
Here the localized position $y_0$ is uniquely determined by $k_x$ via
\begin{align}
 y_0
 =
\hbar k_x/|e|B.
 \label{eq:position_momentum}
\end{align}
When the circumference of the cylinder is $L$, the wave number is discretized as $k_x=2\pi n/L\,(n\in\mathbb{Z})$ and accordingly, localized positions of the Landau levels take discrete values with the interval $\delta y = 2\pi\hbar/|e|BL$.

In Laughlin's argument,
one considers an adiabatic process in which 
a magnetic flux quantum $\Phi_0=2\pi\hbar/|e|$
is threaded through the cylinder.
Corresponding change in the vector potential is $\bm{A}\to\bm{A}+(2\pi\hbar/|e|L,0)$.
If an electron state is coherent along a closed loop in the $x$ direction, a magnetic flux induces a phase shift by $\psi\to e^{2\pi ix/L}\psi$ 
that results in a shift of the momentum by $k_x\to k_x+2\pi/L$.
According to (\ref{eq:position_momentum}), 
the momentum shift is accompanied by an adiabatic 
shift of the electron position from $y_0$ to $y_0+\delta y$.
Such an adiabatic motion of electrons 
forced by threading a magnetic flux 
is the key property in the bulk argument.
Note that electrons without coherence undergo 
trivial changes in their phase factors without any real-space motions.
When the length of the cylinder is infinite, 
or when two boundaries of the cylinder are connected to make a 2-torus, 
all coherent electrons are shifted to their neighboring positions 
by one magnetic flux quantum $2\pi\hbar/|e|$, 
and thus totally the electron state turns back to the original state.

When the electric field is applied in the $y$ direction, 
an electron localized at $y_0$ gains  
an energy by $\delta E= eE_y\delta y$
during a shift to $y_0+\delta y$.
The charge current is given by 
$e\bm{j}=\partial \epsilon/\partial \bm{A}$,
where $\epsilon$ is the electron energy per unit area.
When electrons fill up to the $\nu$th Landau level, the current density is evaluated as
\begin{align}
 ej_x
 =
 \frac{1}{L\delta y}
 \frac{\partial E}{\partial A_x}
 \simeq
 \frac{1}{\delta y}
 \frac{\delta E}{\Phi_0}
 =
 \nu
 \frac{e^2}{2\pi\hbar}
 (-E_y),
 \label{eq:laughlinoriginal}
\end{align}
since all filled Landau levels contribute equally to the Hall current.
In (\ref{eq:laughlinoriginal}),
a differential is approximated by a difference in the second equality.

\section{Boundary argument for the quantum Hall effect}
\label{sec:boundary_qhe}

In this section, 
we revisit Laughlin's argument for the quantum Hall effect
in terms of the $c=1$ chiral boundary theory
\begin{align}
 S
 =
 \int
 d^2x\,
 \bar{\psi}
 i\hbar
 \left(
 \partial_{t}
 +
 \partial_x
 \right)
 \psi,
 \label{eq:chiral_fermion}
\end{align}
 and its intrinsic anomalies.
Here and henceforth we set the Fermi velocity as $v_{\text{F}}=1$.
The chiral boundary theories cannot exist 
as an isolated $(1+1)$-dimensional system, 
and are always accompanied with the higher-dimensional bulk.
The quantum anomaly in the U(1) gauge symmetry and the resulting
breakdown of the charge conservation are peculiarities in such systems, 
and are shown to have a close connection 
with the quantum Hall effect in the bulk.
[Here, we consider the sharp boundary with thickness much shorter
than the magnetic length $l$ to rule out the possibility of edge
reconstruction\cite{chamon94}.
While the subsequent calculations are presented in terms of the simplest
edge theory \eqref{eq:chiral_fermion},
the edge reconstruction is not expected to change the quantum anomaly (the
chiral central charge).]

\subsection{From perturbative U(1) gauge anomaly}
\label{eq:boundary_qhe_perturbative}

\begin{figure}[t]
 \centering
 \includegraphics[width=84mm]{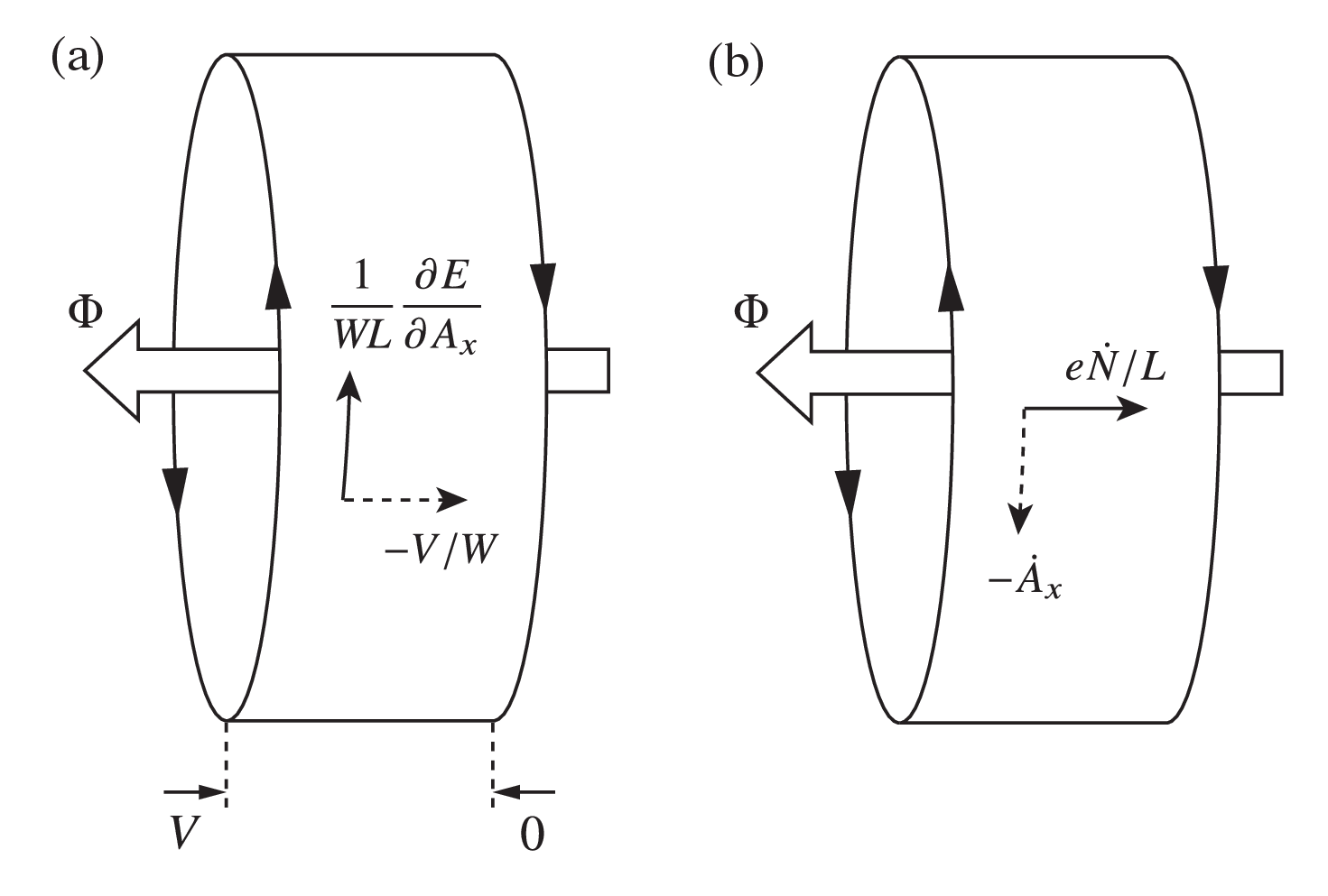}
 \caption{The cylindrical geometry for the boundary picture of Laughlin's argument. On the cylindrical surface, solid arrows represent electric currents and dashed arrows are the electric field. (a) Electric Hall current in the $x$ direction is induced by an applied voltage $V$. (b) Charge pumping between boundaries as an electric Hall current in the $y$ direction is induced by a temporal change of the magnetic flux $\Phi$.\label{fig:laughlin_electric}}
\end{figure}

\subsubsection{Charge pumping and anomaly}

Consider electrons forming a $\nu=1$ quantum Hall state on the cylindrical geometry [Fig.~\ref{fig:laughlin_electric}(a)].
The axial length and the circumference of the cylinder are $W$ and $L$, respectively.
When a magnetic flux $\Phi_0$ is threaded through the cylinder, 
coherent electrons in the bulk flow adiabatically along the cylinder.
At interfaces between boundaries and the bulk, 
bulk electrons flow into the left boundary, 
and simultaneously, 
bulk electrons are supplied by the right boundary.

When we focus on the two boundaries, 
such a process is interpreted by increase and decrease of the electron numbers in the (1+1)-dimensional electronic systems.
The right-(left-)moving chiral boundary fermion mode
resides on the left(right) boundary.
In the following, we denote ``right'' and ``left'' in the subscript of any physical quantities to represent boundary sides, not the moving directions.
Combining two chiral modes, the boundary action is given by
\begin{align}
 S_{\text{left}+\text{right}}
 =&
 \int
 d^2x\,
 \bar{\psi}
 (-i)
 \gamma^{\mu}
 (\hbar\partial_{\mu}
 -
 ie
 A_{\mu})
 \psi,
\end{align}
where $\psi=(\psi_{\text{left}},\psi_{\text{right}})$, $\bar{\psi}=\psi^{\dagger}\gamma^0$, $\gamma^0=i\sigma^x$, $\gamma^1=\sigma^y$ satisfying $\{\gamma^{\mu},\gamma^{\nu}\}/2=\eta^{\mu\nu}=\text{diag}[-1,+1]$, and $\partial_{\mu}=(\partial_t,\partial_x)$.

As electrons flow into/from the left/right boundary,  
the chiral U(1) particle number conservation is violated. 
This is quantified by the chiral U(1) anomaly\cite{bertlmann96,fujikawa04}
equation
\begin{align}
 \partial_{\mu}j_5^{\mu}
 =
 -
 \frac{e}{\pi\hbar}
 \epsilon^{\mu\nu}
 \partial_{\mu}
 A_{\nu},
 \label{eq:chiral_u1_gauge_anomaly}
\end{align}
where $\mu,\nu=t,x$ and $j_5^{\mu}=j_{\text{left}}^{\mu}-j_{\text{right}}^{\mu}$ is the axial current composed of the particle current on left and right boundaries.
Integrating (\ref{eq:chiral_u1_gauge_anomaly}) over the boundary space, one obtains
\begin{align}
 \dot{N}_{\text{left}}-\dot{N}_{\text{right}}
 =
 -
 \frac{e}{\pi\hbar}
 \dot{\Phi},
\end{align}
where $N_{\text{left(right)}}$ is the total electron number of the left (right) boundary defined by the electron density $j^0_{\text{left(right)}}$, and $\Phi$ is the magnetic flux threaded at the center of the boundary circle.
On the other hand, 
the U(1) gauge symmetry of the combination of left and right boundary electrons 
imposes 
the conservation of the total electron number 
$\partial_{t}(N_{\text{left}}+N_{\text{right}})=0$.
Therefore, through adiabatically threading a magnetic flux, the electron number changes as\footnote{
While we have used the chiral U(1) anomaly to quantify the charge pumping,
we could have used the U(1) gauge anomaly, focusing on a single edge. 
The (covariant but not consistent) U(1) anomaly quantifies the loss/gain of 
the charge for a given edge. 
}
\begin{align}
 \delta N_{\text{left}}
 =
 -
 \delta N_{\text{right}}
 =
 -
 \frac{e}{2\pi\hbar}
 \Phi.
 \label{eq:qh_boundary_streda}
\end{align}

\begin{figure}[t]
 \centering
 \includegraphics[width=68mm]{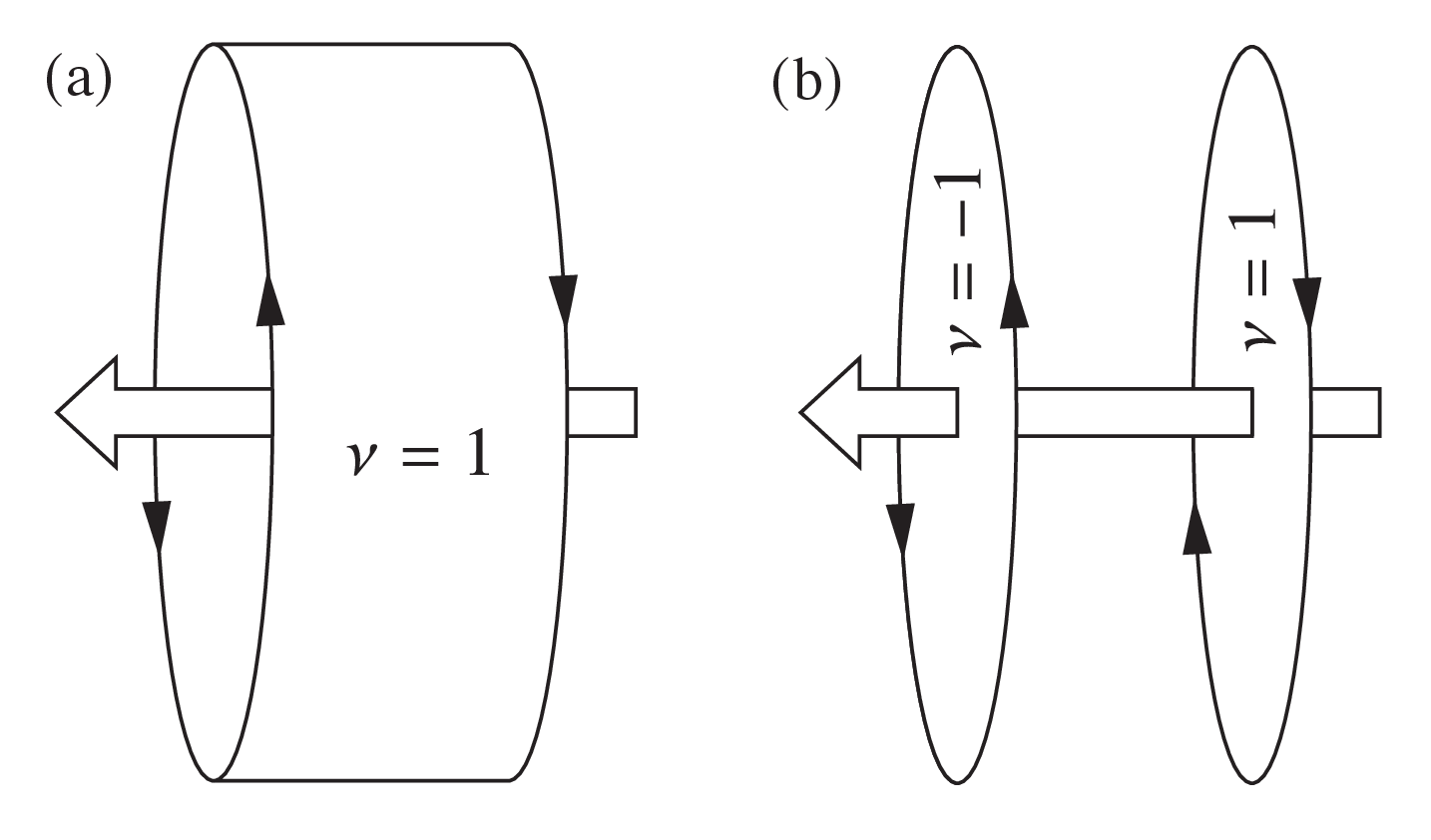}
 \caption{A quantum Hall states on the cylindrical surface have the same boundary electronic modes as those on two disks.\label{fig:geometry}}
\end{figure}

A relation (\ref{eq:qh_boundary_streda}) governing non-conservation of the boundary electron has the same form as the St\v{r}eda formula for the quantum Hall effect\cite{streda82} 
\begin{align}
 \sigma_H=\nu\frac{e^2}{2\pi\hbar}=e\frac{\partial N}{\partial \Phi},
 \label{eq:qh_streda}
\end{align}
with $\nu=-1$ for the left boundary and $\nu=+1$ for the right one, although (\ref{eq:qh_streda}) considers a magnetic flux $\Phi$ that is applied perpendicularly to the two-dimensional electrons, while that in (\ref{eq:qh_boundary_streda}) is threaded through the cylinder.
However, the St\v{r}eda formula (\ref{eq:qh_streda}) and the relation (\ref{eq:qh_boundary_streda}) can be identified as follows, provided that the total electrons number $N$ in (\ref{eq:qh_streda}) is completely due to the chiral boundary modes.
Consider quantum Hall states on two disks $D_{\text{left}}$ and $D_{\text{right}}$ perpendicular to threaded magnetic flux, which have common boundaries with the cylinder as shown in Fig.~\ref{fig:geometry}.
Focusing only on the boundary mode, the chiral boundary modes of the $\nu=1$ quantum Hall state on the cylindrical surface are equivalent to those of the $\nu=-1$ quantum Hall state on $D_{\text{left}}$ and the $\nu=1$ quantum Hall state on $D_{\text{right}}$, where, in the latter geometry, electrons on the cylindrical surface are absent.
This explains the reason why the boundary electrons obey the St\v{r}eda formula.

\subsubsection{The quantized Hall current  
induced by the electrostatic potential}
\label{sec:boundary_qhe_staticpotential}

Let us now relate \eqref{eq:qh_boundary_streda}
to the Hall conductance.
When an magnetic flux $\Phi$ is applied,
the number of electrons on the left boundary at the electric potential $V$ changes by $\delta N_{\text{left}}$
and that on the right boundary at the electric potential $0$ by $\delta N_{\text{right}}$.
The electric potential energy gains by $\delta E_{\text{pot}}=eV\delta N_{\text{left}}$, and, in turn,
the total (kinetic) energy of electrons increases by
\begin{align}
 \delta E
 =
 -eV\delta N_{\text{left}}.
 \label{eq:qh_energy_change}
\end{align}
The electric Hall current is determined by equating 
the energy supplied by applied voltage and 
the interaction energy of the electric current with the vector potential resulting from the threaded magnetic flux $A_x=\Phi/L$.
Thus, using (\ref{eq:qh_boundary_streda}), the electronic current is given by
\begin{align}
 eJ_x
 \equiv
 e\int_0^W dy j_x
 =
 \frac{1}{L}\frac{\partial \delta E}{\partial A_x}
 =
 \frac{e^2}{2\pi\hbar}V.
 \label{eq:quantumhall_boundary}
\end{align}
The above argument can be regarded as a boundary picture of Laughlin's argument on the quantum Hall effect.

Notice that the boundary argument in this subsection cannot tell whether the Hall current flows in the bulk or along the boundary, since it predicts only the total electric Hall current flowing perpendicular to the applied voltage
$
 eJ_x
 \equiv
 e\int_0^W dy j_x
$,
i.e., we have computed the Hall conductance, but not the Hall conductivity.
Provided 
that the electric current is uniformly distributed in the bulk,
we would conclude that the 
electric Hall conductivity is quantized
as in (\ref{eq:laughlinoriginal}),
from (\ref{eq:quantumhall_boundary})
(recalling $E_y=-V/W$).

Alternatively, one can consistently make an argument based on the electric Hall current flowing along the boundary \cite{halperin82, buttiker88}.
This way of describing the Hall current results from a quantization of the boundary electric current
\begin{align}
 e\frac{\partial j_{\text{bdry}}}{\partial \mu}
 =
 \pm\frac{e}{2\pi\hbar},
\end{align}
where $+(-)$ is for the right-(left-)moving chiral fermion.
Then the Hall current is calculated solely by a summation of the boundary current on left and right boundaries as
\begin{align}
 J_x
 =
 j_{\text{left}}+j_{\text{right}}
 =
 \frac{\mu_{\text{left}}-\mu_{\text{right}}}{2\pi\hbar}
 =
 \frac{e}{2\pi\hbar}V,
 \label{eq:quanumhall_fromboundarycurrent}
\end{align}
which is equivalent to (\ref{eq:quantumhall_boundary}).
It should be noted, however, the relation (\ref{eq:quanumhall_fromboundarycurrent}) does not assert that the Hall current is carried only by the chiral boundary modes.
This is because (\ref{eq:quanumhall_fromboundarycurrent}) considers only 
the difference of the electric currents on two boundaries flowing in opposite directions.
 An absolute value of the boundary electric current is not well-defined for the (1+1)-dimensional Dirac system: it depends on the momentum cutoff $\Lambda$ as
\begin{align}
 j_{\text{left/right}}
 \equiv
 \frac{\langle \hat{v}\rangle}{L}
 \simeq
 \frac{1}{2\pi}
 \int_{\pm\Lambda}^{\pm\mu/\hbar}
 dk
 =
 \pm
 \left(\frac{\mu}{2\pi\hbar}
 -
 \frac{\Lambda}{2\pi}
 \right),
\end{align}
while such a high-frequency regime is not well-defied as the boundary property, and should be attributed to the bulk electronic states.
Therefore, the boundary argument in this section does not provide any information about the distribution of the electric Hall current.

\subsubsection{The quantized Hall current 
induced by the time-dependent magnetic flux}
\label{sec:boundary_qhe_tdflux}

Another point to be mentioned is that one can also regard an adiabatic electron transfer between two edges as the electric Hall current flowing in the $y$ direction [Fig.~\ref{fig:laughlin_electric}(b)], which flows perpendicularly to the Hall current (\ref{eq:quantumhall_boundary}) flowing in the $x$ direction [Fig.~\ref{fig:laughlin_electric}(a)].
The Hall current in this case is induced by a temporal change of the magnetic flux which works as the electric field in the $x$ direction: $E_x=-\dot{A_x}=-\dot{\Phi}/L$.
The voltage between two boundaries is absent in this case ($V=0$), and therefore the bulk electronic states are still in equilibrium during threading the magnetic flux.
The electric current density in the bulk is determined by imposing electron number conservation $Lj_y-\delta\dot{N}_{\text{left}}=0$ at the left boundary.
By using (\ref{eq:qh_boundary_streda}), the charge current is related to the electric field as
\begin{align}
 ej_y
 =
 \frac{e\delta\dot{N}_{\text{left}}}{L}
 =
 -\frac{e^2}{2\pi\hbar}
 \frac{\dot{\Phi}}{L}
 =
 \frac{e^2}{2\pi\hbar}
 E_x
\end{align}
which is equivalent to (\ref{eq:quantumhall_boundary}) by rotating $\pi/2$ in the $x$-$y$ plane.

\subsection{From global U(1) gauge anomaly}
\label{From global U(1) gauge anomaly}

The charge pumping relation (\ref{eq:qh_boundary_streda}) 
derived from the chiral U(1) gauge anomaly
can also be derived from another type of anomaly that occurs in the (1+1)-dimensional chiral fermionic system, that is, the global U(1) gauge anomaly.

The U(1) gauge symmetry of the fermionic system refers to invariance 
under a U(1) gauge transformation
\begin{align}
 \psi(x)
 \to
 \psi'(x)
 =
 e^{2\pi i a(x)}
 \psi(x).
\end{align}
In order for the fermionic system on a closed one-dimensional space of the circumference $L$ to be invariant, 
the U(1) gauge transformation must preserve the boundary condition in the spatial direction, 
which is dictated as $a(L)-a(0)\in\mathbb{Z}$.
U(1) gauge transformations satisfying $a(L)-a(0)=0$ can be continuously deformed to the identity transformation ($a(x)=0$), referred to as infinitesimal or small U(1) gauge transformations.
The relation (\ref{eq:qh_boundary_streda}) is a consequence 
of an anomaly regarding transformations of this class, 
which is referred to as the perturbative anomaly.
On the other hand, 
when $a(L)-a(0)=n$ is a nonzero integer, such transformations cannot be continuously deformed to the identity transformation, and are referred to as 
large U(1) gauge transformations.
Threading a magnetic flux
is equivalent to a large U(1) gauge transformation,
when the magnetic flux is an integer multiple of the flux quantum, $a(L)-a(0)=\Phi/\Phi_0$.

Laughlin's original argument on the quantum Hall state considers 
a large U(1) gauge transformation for the \textit{bulk} electronic states 
induced by threading a magnetic flux quantum.
We review the consequence of the same transformation 
on the \textit{boundary} theories\cite{ryu12,witten16}.
Consider the (1+1)-dimensional right-moving chiral fermion on a circle with the circumference $L$ given by
\begin{align}
 H
 =
 \int_0^L dx
 \psi^{\dagger}(x)
 (-i)\left(
 \hbar\partial_x
 -i
 eA_x
 \right)
 \psi(x),
 \label{eq:hamiltonian_flux}
\end{align}
where the electromagnetic vector potential is induced by the magnetic flux $\Phi$ threaded into the center of the circle, and is related via $A_x=\Phi/L$.
We incorporate the effect of $A_x$ 
as a twisted boundary condition in the $x$ direction. 
More generically, 
we consider 
the Hamiltonian
\begin{align}
 H
 =
 \int_0^L
 dx
 \psi^{\dagger}
 (-i\hbar\partial_x)
 \psi
 \label{eq:leftmoving hamiltonian}
\end{align} 
together with 
a twisted boundary condition
in time as well:
\begin{align}
 &\psi(t,x+L)=e^{2\pi i(a-1/2)}\psi(t,x), 
 \label{eq:bc spatial} \\
 &\psi(t+\hbar \beta,x)=
 e^{2\pi i(b-1/2)}\psi(t,x).
 \label{eq:bc temporal} 
\end{align}
Parameters $a,b$ play the role of 
the spatial and temporal flux,
specifically, as
$a-1/2=\Phi/\Phi_0$.
In the canonical formalism,
the temporal twist is realized by
an operation of
$\exp (2\pi i b N)$,
where $N$ is the total fermion number operator
\begin{align}
 N
 =
 \int_0^L dx
 \psi^{\dagger}
 \psi.
\end{align}

Observe that the classical system,
as defined by the Hamiltonian (action) and 
the boundary conditions \eqref{eq:bc spatial} and
 \eqref{eq:bc temporal},
is invariant under 
$a\to a+1$
and 
$b\to b+1$. 
This large gauge invariance, however, 
may be lost once we quantize the theory. 
In particular, the partition function 
may acquire an anomalous phase factor
(= global U(1) gauge anomaly)
under $a\to a+1$ and $b\to b+1$. 

The partition function of the (1+1)-dimensional chiral complex fermion
(\ref{eq:leftmoving hamiltonian})
with the twisted boundary conditions 
can be explicitly computed as follows.
The fermion field operator is expanded by wave functions satisfying (\ref{eq:bc spatial}) as
\begin{align}
 \psi(x)
 =
 \sum_{r\in\mathbb{Z}+a-1/2}
 e^{2\pi ir x/L}
 \psi_r,
\end{align}
and the ground state is defined by filling all negative-energy states.
When $a\in[-1/2,1/2)$,
normal-ordering of the Hamiltonian and the fermion number operator gives
\begin{align}
 H
 &=
 \frac{2\pi \hbar}{L}
 \left[
 \sum_{r\in\mathbb{Z}+a-1/2}
 r:\psi_r^{\dagger}\psi_r:
 -\frac{1}{24}
 +\frac{a^2}{2}
 \right], \\ 
 N
 &=
 \sum_{r\in\mathbb{Z}+a-1/2}
 :\psi_r^{\dagger}\psi_r:
 +
 a,
\end{align}
where extra terms are resulting from the normal-ordering regularized by the Riemann zeta function.
Recall that the partition function of the (1+1)-dimensional chiral fermion without twisting ($a=1/2, b=0$) is given by tracing $e^{-\beta H}$ over the Hilbert space satisfying the periodic boundary condition $\psi(x+L)=\psi(x)$.
The partition function in the present case is given by
\begin{align}
 Z_{[a,b]}
 &\equiv
 \text{Tr}
 \left[
 e^{-\beta H}
 e^{2\pi ibN}
 \right] \notag\\
 &=
 q^{-1/24+a^2/2}e^{2\pi iab} \notag\\
 &\quad\times
 \prod_{n\in \mathbb{N}}
 \left(
 1+q^{n-1/2+a}e^{2\pi ib}
 \right)
 \left(
 1+q^{n-1/2-a}e^{-2\pi ib}
 \right).
 \label{eq:partition function qhe}
\end{align}
where the tracing refers to the boundary condition (\ref{eq:bc spatial}) 
and $q=\exp(-2\pi\hbar\beta/L)$.

By inspection,
one verifies
\begin{align}
	Z_{[a,b]} = Z_{[a+1, b]}
	=
	e^{-2\pi i a} Z_{[a,b+1]}
\end{align}
and hence there is a global U(1) gauge anomaly. 
From the anomaly,
we can read off the charge pumping formula.  
We normalize the particle number
such that the ground state particle number at $a=0$ ($\Phi=-\Phi_0/2$)
as 0. 
At $a=0$, by changing the chemical potential $b\to b+1$ 
one does not earn any phase. 
On the other hand, at $a\neq 0$, 
the partition function acquires a non-zero phase factor. 
This phase is indicative of the change of the ground state fermion number 
as compared to the fermion number at $a=0$. 
Since the free energy changes by $\delta F=-2\pi ia/\beta$
during the change of the chemical potential $\delta \mu=2\pi i/\beta$, the particle number is evaluated as $N=-\delta F/\delta \mu = a$.
(Note that, from \eqref{eq:partition function qhe}, the ``imaginary'' chemical potential is identified as $\beta\mu=2\pi i b$.) Then,
\begin{align}
 \frac{\partial N}{\partial \Phi}
 =
 \frac{1}{\Phi_0}
 \frac{\partial N}{\partial a}
 =
 \frac{|e|}{2\pi \hbar},
 \label{eq:globalu1gaugeanomaly}
\end{align}
which is equivalent to the consequence of the perturbative U(1) gauge anomaly (\ref{eq:qh_boundary_streda}), although broken symmetries are distinct.

To give a more microscopic view on the global U(1) gauge anomaly,
let us follow the spectrum of the Hamiltonian
\eqref{eq:hamiltonian_flux}
as we change the magnetic flux adiabatically. 
Under the periodic boundary condition, the eigenfunction is $\phi_n(x)=\exp[2\pi i nx/L]/\sqrt{L}\,(n\in\mathbb{Z})$, and the corresponding eigenenergy is 
\begin{align}
 \epsilon_n(\Phi)
 =
 \frac{2\pi n\hbar}{L}
 -
 \frac{e\Phi}{L}.
\end{align}
After a magnetic flux quantum $\Phi_0=2\pi\hbar/|e|$ is threaded, the energy spectra turn back to the original ones by shifting each energy level to the adjacent one ($\epsilon_n\to\epsilon_{n+1}$).
This implies that the large U(1) gauge transformation leaves the whole electronic energy spectra invariant.
However, following gradual change of the energy spectra through threading the magnetic flux, the ground state property changes.

Quantizing the fermion by introducing the anticommutation relation $\{\psi(x),\psi^{\dagger}(x')\}=\delta(x-x')$, and expanding the fermion operator by the eigenmodes $\psi(x)=\sum_n\phi_n(x)c_n$, the Hamiltonian is rewritten as
\begin{align}
 H
 =
 \sum_n
 \epsilon_n(\Phi)
 c_n^{\dagger}c_n.
\end{align}
If the magnetic flux initially lies in the range $\Phi\in(-\Phi_0,0)$, the eigenenergy is positive for $n > 0$ and negative for $n \le 0$. 
The ground state $|0\rangle_{\Phi}$ is made by filling all states with negative eigenenergies, thus the Fermi level lies between $\epsilon_0(\Phi)$ and $\epsilon_1(\Phi)$.
After threading a magnetic flux quantum $\Phi_0$, each energy level is shifted as $\epsilon_n(\Phi+\Phi_0)=\epsilon_{n+1}(\Phi)$, and thus
the new Fermi level lies between $\epsilon_{1}(\Phi)$ and $\epsilon_2(\Phi)$.
While the energy spectra are invariant through the magnetic flux change $\Phi\to\Phi+\Phi_0$, the number of electrons in the ground state changes by unity $\delta N=1$, since a filled energy level with eigenenergy $\epsilon_{0}(\Phi+\Phi_0)(=\epsilon_{1}(\Phi))$ goes above the original Fermi level.
Let the electron number change be a continuous function of the threaded magnetic flux, the above relation, again, leads to
\eqref{eq:globalu1gaugeanomaly}.

As shown above, the global U(1) anomaly counts the number of electronic energy levels that traverse the Fermi level during the large U(1) gauge transformation.
Within this process, only energy levels close to the Fermi level are concerned.
Therefore, as long as the transformation leaves the electronic system invariant at the classical level, the quantized number of traversed energy levels would be unaffected even after small perturbations are added.

\section{Boundary argument for the quantized thermal Hall effect}
\label{sec:boundary_the}

As seen in the previous section, 
the quantum Hall effect can be explained by anomalies 
of the (1+1)-dimensional chiral boundary theory.
The broken symmetries in these arguments 
are the invariance under 
infinitesimal and large U(1) gauge transformations.
Here, we extend this boundary argument to the case of the quantized thermal Hall effect.
With the help of the St\v{r}eda formula for the quantized thermal Hall effect\cite{nomura12}
\begin{align}
 \kappa_H
 =
 c\frac{\pi k_{\text{B}}^2T}{6\hbar}
 =
 \frac{\partial S}{\partial \Phi^\text{g}},
 \label{eq:qth_streda}
\end{align}
the relevant symmetry is described by a spacetime transformation given in terms of gravity.

\subsection{Modular transformation}

Consider the (1+1)-dimensional system under a static gravitational field $g^{\mu\nu}$.
The St\v{r}eda formula for the quantized thermal Hall effect (\ref{eq:qth_streda}) describes an entropy change induced by the gravitomagnetic flux, which is the gravitational counterpart of the magnetic flux defined by $\Phi^\text{g}=A^\text{g}_xL$.
The gravitomagnetic vector potential $A^\text{g}_x$ is defined by the line element of the Minkowski spacetime  
\begin{align}
 ds^2
 =
 -(dt+A^\text{g}_x dx)^2
 +
 dx^2.
 \label{eq:euclidean_spacetime}
\end{align}
By the Wick rotation, the line element of the Euclidean spacetime is given by
\begin{align}
 ds^2
 =
 (dt^\text{E}+A^\text{E}_x dx)^2
 +
 dx^2,
 \label{eq:euclidean_spacetime}
\end{align} 
where $t^\text{E}=it$ is the imaginary time, and $A_x^\text{E}=iA_x^\text{g}$ is the gravitomagnetic vector potential in the Euclidean spacetime.
In the following, the symbol $t$ is used as the imaginary time in place of $t^\text{E}$ for convenience.
In the finite-temperature formalism, boundaries of the temporal direction are periodically identified 
with the period $\hbar\beta=\hbar/(k_{\text{B}}T)$.
When the space direction has also the periodic boundary by the period $L$, the spacetime is a 2-torus.

\begin{figure}[t]
 \centering
 \includegraphics[width=34mm]{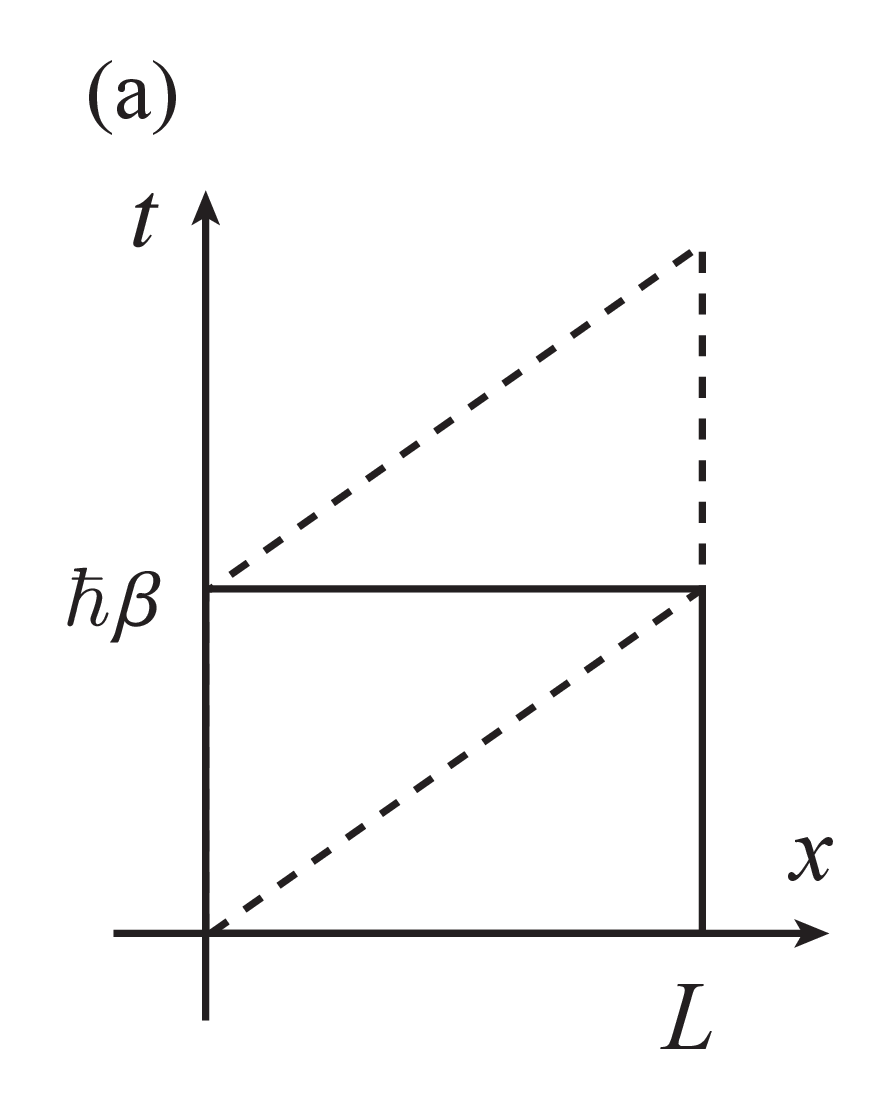}\\
 \vspace{5mm}
 \includegraphics[width=88mm]{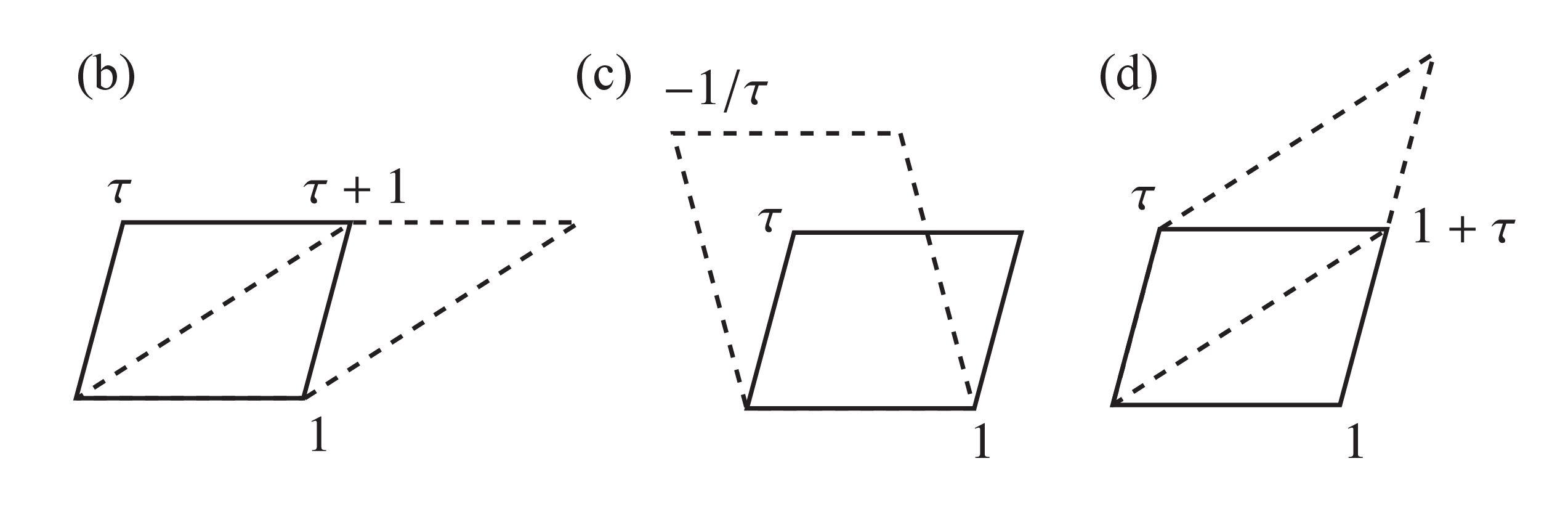}
 \caption{(a) A modular transformation of the spacetime 2-torus induced by threading a gravitomagnetic flux $\Phi^\text{E}=\hbar\beta$. A rectangular 2-torus (solid line) is transformed to a sheared-rectangular 2-torus (dashed line).
 Boundaries on left and right, and those on top and bottom are identified, respectively.
 The modular group is generated by (b) $T:\tau\to\tau+1$, and (c) $S:\tau\to -1/\tau$.
 (d) Another basic transformation can be composed by $TST:\tau\to\tau/(1+\tau)$.  \label{fig:modular}}
\end{figure}

Provided that the gravitomagnetic vector potential $A^\text{E}_x$ is static, a transformation from a flat spacetime to the one specified by (\ref{eq:euclidean_spacetime}) is given by a diffeomorphism
\begin{align}
 (t,x)
 \to
 (t+\hbar\beta a^\text{E}(x), x),
 \label{eq:spacetime_transformation}
\end{align}
where $a^\text{E}(x)=(\hbar\beta)^{-1}\int_0^x dx'A_x^\text{E}(x')$.
Taking into account the fact that the imaginary time is defined modulo $\hbar\beta$, a transformation satisfying $a^\text{E}(L)-a^\text{E}(0)\in\mathbb{Z}$ leaves the spacetime invariant.
Corresponding gravitomagnetic flux $\Phi^\text{E}$ is an integer multiple of $\hbar\beta$.
A transformation (\ref{eq:spacetime_transformation}) with a nonzero integer $a^\text{E}(L)-a^\text{E}(0)$ cannot be continuously deformed to the identity transformation.
This type of transformations is referred to as  
large diffeomorphism.
Large diffeomorphisms of a torus are
referred to as modular transformations\cite{ginsparg89,francesco97}.
Consider a simplest modular transformation given by $\Phi^\text{E}_0\equiv\hbar\beta$ or $A^\text{E}_x=\Phi^\text{E}_0/L$, and a corresponding transformation
\begin{align}
 (t,x)
 \to
 (t',x')
 =
 (t+\hbar\beta x/L, x).
 \label{eq:modular_transformation}
\end{align}
After this transformation, periodicity of the spacetime 2-torus is altered from an identification
\begin{align}
 (t,x)
 \sim
 (t+\hbar\beta,x)
 \sim
 (t,x+L).
 \label{eq:periodicity_before}
\end{align}
to a new identification\cite{golkar16}
\begin{align}
 (t,x)
 \sim
 (t+\hbar\beta,x)
 \sim
 (t+\hbar\beta,x+L),
 \label{eq:periodicity_after}
\end{align}
which is shown in Fig.~\ref{fig:modular} (a).
The transformation (\ref{eq:modular_transformation}) represents a sequence of prescriptions composed of, cutting the spacetime torus by a loop along the temporal direction, twisting one of the edges by $\hbar\beta$, and gluing two edges to make a 2-torus again.
Notice that the unit of the gravitomagnetic flux inducing a modular transformation is $\Phi^\text{g}_0=-i\hbar\beta$, while the unit of the magnetic flux bringing about a large U(1) gauge transformation is the magnetic flux quantum $\Phi_0=2\pi\hbar/|e|$.

Before moving on, 
we briefly review the concept of the modular group\cite{francesco97}.
The spacetime 2-torus is defined by periodicities, and thus is the quotient space of the two-dimensional Euclidean space $\mathbb{R}^2$ by a two-dimensional lattice spanned by two linearly-independent lattice vectors.
When the torus is defined on the complex plane $\mathbb{C}$ by, e.g.~$z=x+it$, the lattice vectors are represented by two complex numbers $\omega_1,\omega_2\in\mathbb{C}$ as
\begin{align}
 (t,x)
 \sim
 (t+\text{Im}[\omega_{1(2)}],x+\text{Re}[\omega_{1(2)}]).
\end{align}
As in the description of crystals, there is an ambiguity in choice of the lattice vectors.
Another set of lattice vectors given by a transformation
\begin{align}
 \begin{pmatrix}
  \omega_2' \\ \omega_1'
 \end{pmatrix}
 =
 \begin{pmatrix}
  a & b \\
  c & d
 \end{pmatrix}
 \begin{pmatrix}
  \omega_2 \\ \omega_1
 \end{pmatrix}
 \label{eq:modulartransformation_latticevector}
\end{align}
satisfying $ad-bc=1\,(a,b,c,d\in\mathbb{Z})$,
spans the same lattice, since the transformation matrix is invertible.
The matrix in (\ref{eq:modulartransformation_latticevector}) leaves the area spanned by two lattice vectors invariant, and forms a group $S\!L(2,\mathbb{Z})$ of $2\times 2$ integer-valued matrices with unit determinant.

Thanks to the conformal invariance that the linearized form of the gapless boundary fermion (\ref{eq:chiral_fermion}) possesses, physical properties on a torus should be invariant up to scaling, and thus be dependent only on the ratio of two periods $\tau=\omega_2/\omega_1$, which is referred to as the modular parameter.
Redefinition of lattice vectors (\ref{eq:modulartransformation_latticevector}) transforms the modular parameter as
\begin{align}
 \tau'
 =
 \frac{a\tau+b}{c\tau+d},
\end{align}
which forms a group $PS\!L(2,\mathbb{Z})=S\!L(2,\mathbb{Z})/\mathbb{Z}_2$, referred to as the modular group.
Here $\mathbb{Z}_2$ in the modular group is due to the fact that inverting the signs of $a,b,c,d$ leaves the transformation unchanged.
The modular group is known to be generated by two operations $T: \tau\to\tau+1$ and $S:\tau\to -1/\tau$ [Fig.~\ref{fig:modular} (b) and (c)].

Here, we apply the above framework to our situation.
The lattice vectors of the rectangular spacetime torus (\ref{eq:periodicity_before}) are assigned as $\omega_1=L$ and $\omega_2=i\hbar\beta$, and corresponding lattice vectors of the sheared rectangular spacetime torus (\ref{eq:periodicity_after}) are $\omega_1=L+i\hbar\beta$ and $\omega_2=i\hbar\beta$.
Defining the ratio of spatial and temporal periods by $\alpha=\hbar\beta/L$, the modular parameter is changed from $\tau=i\alpha$ to $\tau'=i\alpha/(1+i\alpha)$ during the gravitomagnetic flux $\Phi^{\text{E}}_0$ is threaded.
This process is a modular transformation given by $TST:\tau\to\tau/(1+\tau)$ [Fig.~\ref{fig:modular} (d)].

Notice that, in the above context, we have encoded the gravitomagnetic flux into the change of the lattice vectors that span the spacetime torus, not into the change of the metric with which the fermionic kinetic action is defined.
These two interpretations are equivalent, at least, when the gravitomagnetic flux is uniform in the whole spacetime (see for details in appendix \ref{sec:equivalence}).
With this in mind, we study, throughout this paper, the fermionic action on the flat spacetime under the boundary condition specified by the threaded magnetic and gravitomagnetic fluxes.

\subsection{Free energy pumping and global diffeomorphism anomaly}

In this section, the breakdown of the modular invariance,
that is, the global diffeomorphism
anomaly\cite{witten85,ryu12,witten16,golkar16},
of the (1+1)-dimensional edge theory of the quantum Hall systems is reviewed,
and is shown to account for the quantized thermal Hall effect.

For our calculation of the global diffeomorphism anomaly, 
we again employ the chiral massless Dirac fermion theory (\ref{eq:leftmoving hamiltonian}).
It should be stressed that this theory (\ref{eq:leftmoving hamiltonian}) enjoys
an exact conformal (and/or Lorentz) symmetry,
which makes the following calculations rather transparent.
In contrast, the (realistic) edge theory of the quantum Hall boundary
realizes the conformal symmetry only approximately at low energies.
Our rationale of assuming the exact conformal symmetry is
that we focus on the renormalization group fixed point,
which, irrespective of microscopic details, is described by a scale invariant field theory.
For edge theories which are not quite at a renormalization group fixed point,
we invoke the usual 't Hooft anomaly matching,
i.e., 
the calculation of quantum anomalies
should not depend on what energy/length scale is chosen for the calculation.
This should be contrasted with our calculation of the large U(1) gauge anomaly
and the quantized Hall conductance:
The large U(1) gauge invariance is an exact symmetry of the system at all
scales. 
On the other hand, 
in the thermal/gravitational case, at least technically, our calculation of the global
gravitational anomaly (presented below) relies on an emergent conformal symmetry
at low energies. 
We leave it as a future problem 
whether or not the reliance on the conformal symmetry
can be relaxed or completely removed. 
(See, however, Ref.\ \onlinecite{kitaev06}, where it was attempted to give
the definition of the chiral central charge without assuming conformal symmetry.)

The global diffeomorphism anomaly can be read off from the partition function.
In addition to the modular parameter $\tau$ that characterizes the base spacetime manifold, one needs to specify the boundary condition of the fermion defined on it.
The boundary condition is, in general, defined for two periods by
\begin{align}
 &\psi(t+\text{Im}[\omega_1],x+\text{Re}[\omega_1])=e^{2\pi i(a-1/2)}\psi(t,x), 
 \label{eq:bc_tauab1} \\
 &\psi(t+\text{Im}[\omega_2],x+\text{Re}[\omega_2])=e^{2\pi i(b-1/2)}\psi(t,x).
 \label{eq:bc_tauab2} 
\end{align}
The boundary conditions for the fermion on the spacetime torus without the gravitomagnetic flux (\ref{eq:periodicity_before}) is given by
\begin{align}
 \begin{aligned}
 &\psi(t,x+L)=\psi(t,x),
 \\
 &\psi(t+\hbar\beta,x)=-\psi(t,x),
 \end{aligned}
 \label{eq:bc_before}
\end{align}
which corresponds to $\tau=i\alpha$ and $[a,b]=[\frac{1}{2},0]$.
On the other hand, the boundary condition on a torus with the gravitomagnetic flux $\Phi^\text{E}_0$ specified by (\ref{eq:periodicity_after}) is 
\begin{align}
 \begin{aligned}
 &\psi(t+\hbar\beta,x+L)=-\psi(t,x),
  \\
 &\psi(t+\hbar\beta,x)=-\psi(t,x),
 \end{aligned}
 \label{eq:bc_after}
\end{align}
which corresponds to $\tau=i\alpha/(1+i\alpha)$ and $[a,b]=[0,0]$.
If the fermionic system is invariant under the modular transformation, the partition function should be unchanged during the transformation.
This is not true for the present case since there is an anomaly regarding the modular invariance.

The partition function of the (1+1)-dimensional chiral complex fermion (\ref{eq:leftmoving hamiltonian}) with the boundary condition specified by the modular parameter $\tau=\tau_1+i\tau_2\alpha$ and $[a,b]$ is calculated, 
in much the same way as in Sec. \ref{From global U(1) gauge anomaly}:
\begin{align}
 &Z_{[a,b]}(\tau)
 \equiv
 \text{Tr}
 \left[
 e^{-\tau_2\beta H}
 e^{i\tau_1LP/\hbar}
 e^{2\pi ibN}
 \right] \notag\\
 &=
 q^{-1/24+a^2/2}e^{2\pi iab} \notag\\
 &\quad\times
 \prod_{n\in \mathbb{N}}
 \left(
 1+q^{n-1/2+a}e^{2\pi ib}
 \right)
 \left(
 1+q^{n-1/2-a}e^{-2\pi ib}
 \right),
\end{align}
where $P=H$ (we have set the Fermi velocity as $v_{\text{F}}=1$) and $q=e^{2\pi i\tau}$.
Under  the modular transformation $TST:\tau\to\tau/(1+\tau)$, the partition function of the boundary fermion is transformed as\cite{ginsparg89,francesco97}
\begin{align}
 Z_{[\frac{1}{2},0]}(i\alpha)
 \to&
 Z_{[0,0]}(i\alpha/(1+i\alpha)) \notag\\
 &=
 e^{-i\pi/12} 
 Z_{[0,\frac{1}{2}]}(-1/(1+i\alpha)) \notag\\
 &=
 e^{-i\pi/12} 
 Z_{[\frac{1}{2},0]}(1+i\alpha) \notag\\
 &=
 e^{i\pi/12} 
 Z_{[\frac{1}{2},0]}(i\alpha).
\end{align}
A contribution due to the global diffeomorphism anomaly appears as an extra phase factor $e^{i\pi/12}$.
Therefore an extra \textit{imaginary} free energy $\delta F=-i\pi/12\beta$ is generated during this process.
Since a \textit{real} gravitomagnetic flux $\Phi^\text{E}_0=\hbar\beta$ in the Euclidean spacetime corresponds to an \textit{imaginary} gravitomagnetic flux $\Phi^\text{g}_0=-i\hbar\beta$ in the Minkowski spacetime, a free energy change induced by the gravitomagnetic flux is formulated as
\begin{align}
 \frac{\partial F}{\partial \Phi^\text{g}}
 \simeq
 \frac{\delta F}{\Phi^\text{g}_0}
 =
 \frac{\pi k_{\text{B}}^2T^2}{12\hbar},
 \label{eq:globaldiffeomorphismanomaly}
\end{align}
where in the first equality, a differential is approximately given by a difference as in the case of the global U(1) gauge anomaly (\ref{eq:globalu1gaugeanomaly}).
An indication of the relation (\ref{eq:globaldiffeomorphismanomaly}) is that the (1+1)-dimensional gapless fermionic system loses or gains free energy depending on its central charge, by threading the gravitomagnetic flux into the one-dimensional space loop.

The free energy (\ref{eq:globaldiffeomorphismanomaly}) 
has been derived and discussed in the context of 
the anomaly-related transport phenomena
\footnote{See for example,
R. Loganayagam and P. Sur{\'o}wka, J. High Energy Phys. \textbf{4}, 097 (2012),
K. Jensen, R. Loganayagam, and A. Yarom, J. High Energy Phys. \textbf{2}, 088 (2013).
In these works,
the anomaly-related finite-temperature transport coefficients and free energy in even dimensions are discussed,
and the so-called ``replacement rule'' connecting the free energy and 
the anomaly polynomials is proposed, in which 
the field strength 
and the Riemann curvature is ``replaced'' by 
the (chiral) chemical potential and the temperature.}.
In particular, 
Golkar and Sethi \cite{golkar16}
discussed
the free energy (\ref{eq:globaldiffeomorphismanomaly}) 
by using the global gravitational anomaly.
(The same free energy was also obtained in 
Ref.\ \onlinecite{nakai16} -- see discussion below.)
It should be noted however that 
this method of determining an effective free energy 
from the global anomaly 
suffers from an ambiguity.
The free energy change can be determined only up
to an integer multiple of $2\pi$,
\begin{align}
\delta F=(-i/\beta)(\pi/12+2\pi n)
\quad (n\in\mathbb{Z}), 
\end{align}
since the logarithm of the extra phase factor $e^{i\pi/12}$ can be determined up to an integer multiple of $2\pi i$ \cite{chowdhury16}.
Nevertheless,
the ambiguity can be removed 
by requiring
the consistency with 
the perturbative gravitational anomaly, and 
the boundary thermal conductivity \cite{cappelli02,chowdhury16}
leading to
the free energy (\ref{eq:globaldiffeomorphismanomaly}).

Observe 
the same ambiguity does exist 
for the case of the global U(1) gauge anomaly:
The global anomaly (the anomalous phase 
acquired by the partition function
under large U(1) gauge transformations)
is determined only up to an integer multiple of $2\pi$.
Once again, 
matching the global anomaly with
the perturbative U(1) gauge anomaly 
removes the ambiguity. 
It should be also noted that, for the case
of the global U(1) gauge anomaly,
the situation is slightly better as there are 
two compact adiabatic parameters, $a$ and $b$, 
that we can change. 
While the anomalous phase 
$\exp(2\pi ia)$ under $b\to b+1$ 
has an ambiguity, 
demanding that the phase is a continuous function
of $a$, one can read off the Hall conductance 
from 
the derivative of 
$\ln [\exp 2\pi i a]$
with respect to $a$,
which is free from the ambiguity.
On the other hand, for the gravitational case, 
we have only one compact variable $\tau$.
We thus need to resort on
consistency with the perturbative gravitational anomaly
to fix the ambiguity.

If we need to fix the ambiguity 
with the help of the perturbative anomaly,
one may wonder why we need to rely on the global anomaly
in the first place. 
However, as noted previously
\cite{stone12}, 
deriving the thermal response by using the
perturbative gravitational anomaly is not obvious, 
as one needs to relate the gravitational response 
to the thermal response by using
Luttinger's trick\cite{luttinger64}. 
On the other hand, 
as we will demonstrate in the following,
the thermal response appears
more naturally when we consider the global diffeomorphism anomaly.

A direct consequence of (\ref{eq:globaldiffeomorphismanomaly}) is the St\v{r}eda formula for the quantized thermal Hall effect.
Using a thermodynamic relation $\delta S=-\partial\delta F/\partial T$, the St\v{r}eda formula is derived as
\begin{align}
 \frac{\delta S}{\Phi^\text{g}_0}
 =
 -\frac{\pi k_{\text{B}}^2T}{6\hbar}
 =
 \kappa_H(c=-1),
 \label{eq:streda_the_anomaly}
\end{align}
where $\kappa_H(c)$ represents the quantized thermal Hall conductivity for the chiral central charge $c$.
(\ref{eq:streda_the_anomaly}) is the St\v{r}eda formula for the quantized thermal Hall effect in the $\nu=-1$ quantum Hall system, and is quite analogous to (\ref{eq:qh_streda}) for the quantum Hall effect led by the U(1) gauge anomaly.

Although the free energy (\ref{eq:globaldiffeomorphismanomaly}) is a functional only of the gravitomagnetic vector potential $A^\text{g}_x$:
\begin{align}
 F[A^\text{g}_x]
 =
 \frac{\pi k_{\text{B}}^2T^2}{12\hbar}
 \int_0^L dx A^\text{g}_x,
 \label{eq:freeenergy_vectorpotential}
\end{align}
one can deduce a form of the free energy when a gravitational potential field $\sigma$ is additionally present.
The metric is given by
\begin{align}
 ds^2
 =
 e^{-2\sigma}
 \left(
 dt
 +
 iA_x^\text{g}dx
 \right)^2
 +
 dx^2.
\end{align}
Thus including a gravitational potential is reduced to changes $\beta\to e^{-\sigma}\beta$ and $A_x^\text{g}\to e^{-\sigma}A_x^\text{g}$.
The global diffeomorphism anomaly in this new metric is read off from the free energy change $\delta F=-i\pi e^{\sigma}/12\beta$ induced by $\Phi^\text{g}=-i\hbar e^{-\sigma}\beta$, which results in the free energy as a functional of $\sigma$ and $A^\text{g}_x$.
Expanding with respect to the gravitational potential as
\begin{align}
 F[\sigma,A^\text{g}_x]
 &=
 \frac{\pi k_{\text{B}}^2T^2}{12\hbar}
 \int_0^L dx e^{2\sigma}A^\text{g}_x \notag\\
 &=
 F[A^\text{g}_x]
 +
 F^{(1)}[\sigma,A^\text{g}_x]
 +
 O(\sigma^2),
\end{align}
the zeroth-order term is given in (\ref{eq:freeenergy_vectorpotential}), while the first-order term
\begin{align}
 F^{(1)}[\sigma,A^\text{g}_x]
 =
 \frac{\pi k_{\text{B}}^2T^2}{6\hbar}
 \int_0^L dx\, \sigma A^\text{g}_x,
\end{align}
is equivalent to the boundary free energy derived by the authors in a previous paper\cite{nakai16}.

\begin{figure}[t]
 \centering
 \includegraphics[width=84mm]{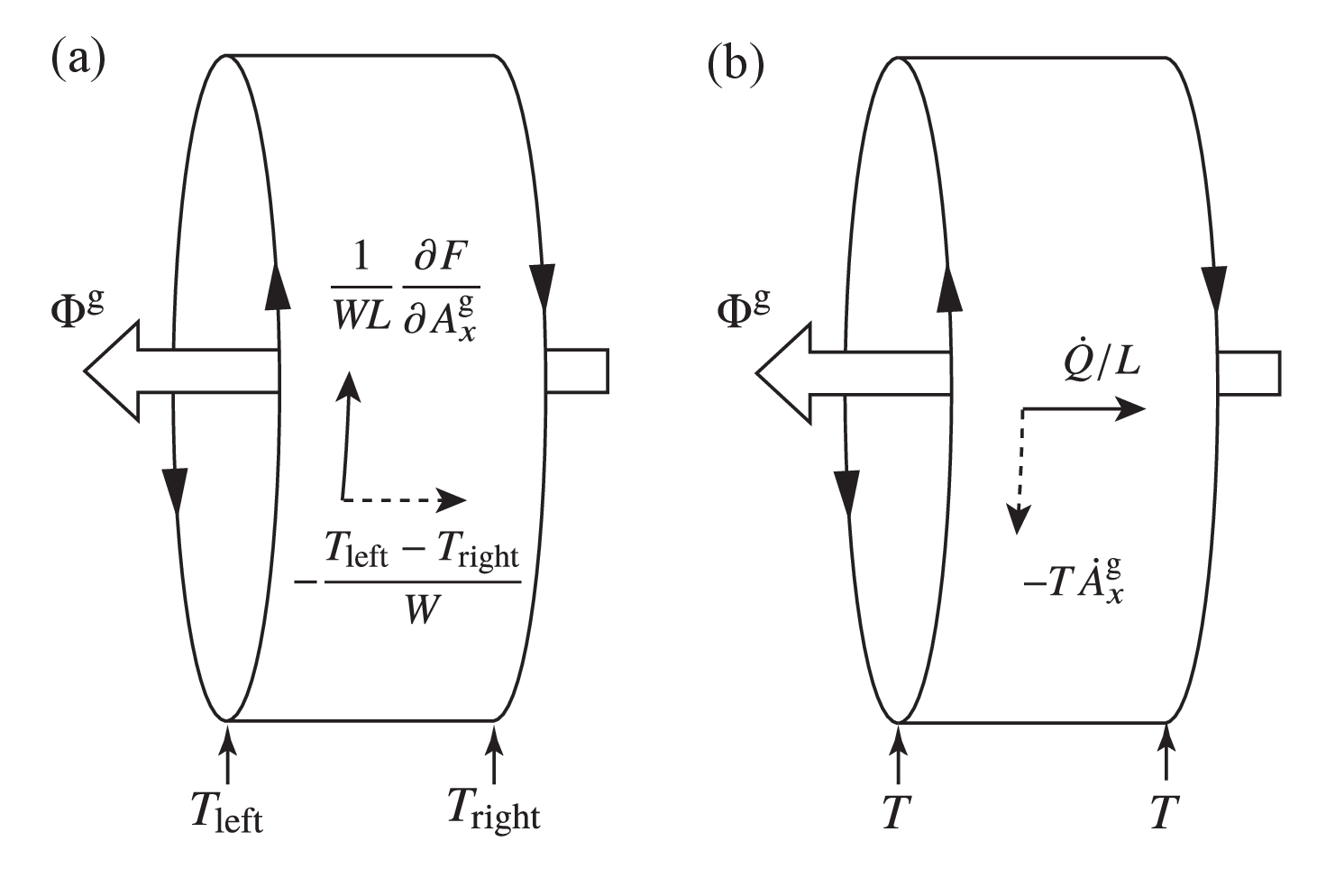}
 \caption{
 A setup for Laughlin's argument on the quantized thermal Hall effect from the boundary theory. 
 On the cylindrical surface, solid arrows represent thermal currents and dashed arrows represent temperature gradients. 
 (a) Left and right boundaries are in contact with heat baths 
and are in thermal equilibrium at temperature $T_{\text{left}}$ and $T_{\text{right}}$, respectively. 
The thermal Hall current is induced by the temperature difference between boundaries.
(b) Two boundaries are in thermal equilibrium at the same temperature $T$. 
A transferred heat between boundaries as a thermal Hall current is induced by a temporal change of the gravitomagnetic flux $\Phi^\text{g}$.\label{fig:laughlin_thermal}}
\end{figure}

\subsection{The quantized thermal Hall current  
induced by the temperature gradient
}
\label{sec:boundary_the_notice}

Now we are ready to extend Laughlin's argument using the relation in the previous subsection (\ref{eq:globaldiffeomorphismanomaly}).
Consider a geometry shown in Fig.~\ref{fig:laughlin_thermal}(a).
The bulk electrons form a quantum Hall state with the Chern number $\nu=1$.
Left and right boundaries are in the thermal equilibrium at temperature $T_{\text{left}}$ and $T_{\text{right}}$, respectively, by contacting them with heat baths.
The quantum Hall state on the cylindrical surface is assumed to have an energy gap much larger than both boundary temperatures so that the electronic excitations are suppressed in the bulk.

The thermal Hall current can be read off by equating the free energy generated at the boundaries 
as a result of the global diffeomorphism anomaly (\ref{eq:globaldiffeomorphismanomaly}), 
and an interaction energy of the thermal current with the gravitomagnetic vector potential induced by the gravitomagnetic flux.
A free energy generated at left and right boundaries is given, respectively, by
\begin{align}
 \delta F_{\text{left}}
 =
 \frac{\pi k_{\text{B}}^2T^2_{\text{left}}}{12\hbar}
 \Phi^\text{g},
 \quad
 \delta F_{\text{right}}
 =
 -\frac{\pi k_{\text{B}}^2T^2_{\text{right}}}{12\hbar}
 \Phi^\text{g},
 \label{eq:freeenergy_leftright}
\end{align}
and thus the change of the total free energy by
\begin{align}
 \delta F
 \equiv
 \delta F_{\text{left}}
 +
 \delta F_{\text{right}}
 =
 \frac{\pi k_{\text{B}}^2}{12\hbar}
 \left(
 T^2_{\text{left}}
 -
 T^2_{\text{right}}
 \right)
 \Phi^\text{g}.
 \label{eq:qth_bulk_freeenergy}
\end{align}
When the temperature difference between two boundaries is sufficiently small compared with boundary temperatures themselves ($|T_{\text{left}}-T_{\text{right}}|\ll T_{\text{left(right)}}$), one obtains 
\begin{align}
 \delta F
 \simeq
 \frac{\pi k_{\text{B}}^2\bar{T}}{6\hbar}
 (T_{\text{left}}-T_{\text{right}})
 \Phi^\text{g},
\end{align}
where $\bar{T}$ is the average temperature between $T_{\text{left}}$ and $T_{\text{right}}$.
The thermal current (energy current) couples to the gravitomagnetic field, and is derived from this free energy as
\begin{align}
 \int^W_0 dy\, j_{x}^T
 =
 \frac{1}{L}
 \left(
 \frac{\partial \delta F}{\partial A_{x}^\text{g}}
 \right)
 =
 \frac{\pi k_{\text{B}}^2\bar{T}}{6\hbar}
 (T_{\text{left}}-T_{\text{right}}),
 \label{eq:qth_boundary}
\end{align}
which is the quantized thermal Hall effect with the thermal Hall  conductance
$\pi k_{\text{B}}^2\bar{T}/6\hbar$ for the Chern number $\nu=1$.
Notice that the boundary argument presented above is free from the fictitious temperature gradient in terms of gravity, that is, Luttinger's trick\cite{luttinger64} using the Tolman-Ehrenfest relation $-T^{-1}\nabla_xT=-\nabla_x \sigma$ by the gravitational potential $\sigma$.

As a result, when two sides of the quantum Hall boundaries contact with heat baths with different temperature, a thermal current flows
parallel to the boundaries, 
and the thermal Hall conductance is quantized by the central charge of the chiral boundary modes, which, in this case, is equivalent to the bulk Chern number.
However it should be noted that the boundary argument cannot tell whether the thermal Hall current flows in the bulk or along the boundary, due to the same reason as we mentioned in Sec.~\ref{sec:boundary_qhe_staticpotential} for the quantum Hall effect.
The relation (\ref{eq:qth_boundary}) tells us about the total thermal Hall current $L^{-1}\partial\delta F/\partial A_{x}^\text{g}$ integrated over the section of the Hall bar geometry.
For example, one can also explain the
thermal Hall effect solely by the boundary thermal current.
The thermal current of the (1+1)-dimensional fermion is evaluated as
\begin{align}
 \frac{\partial j^{T,\text{bdry}}}{\partial T}
 =
(c-\bar{c})
 \frac{\pi k_{\text{B}}^2T}{6\hbar},
 \label{eq:perturbativegravitationalanomaly}
\end{align}
which is related to a perturbative gravitational anomaly \cite{cappelli02}.
Although the relation (\ref{eq:perturbativegravitationalanomaly}) is enough to show the quantized thermal Hall effect when two boundaries have different temperature, we cannot conclude, from this relation, that the thermal Hall current flows only near the boundary.
This is because the absolute value of a thermal current flowing along the boundary cannot be determined.

The boundary argument presented in this section, and the similar one in the previous section for the quantum Hall effect, rely on the presence of the chiral massless fermionic mode on the boundary and the gapful bulk.
The presence of the chiral massless fermion is robust against perturbations including disorders and interaction as long as the bulk energy gap is large enough compared with perturbations.
Furthermore, the boundary mode is robust against perturbations on the boundary due to chirality.
However, unlike the case of the quantum Hall effect where the large U(1) gauge invariance and quantization of electric responses are exact for the chiral boundary modes, the thermal Hall coefficient is not necessarily quantized, in a strict meaning, due to the breakdown of the scale invariance by microscopic details of the model.

\subsection{The quantized thermal Hall current  
induced by the time-dependent gravitomagnetic flux}

Following the discussion of the quantum Hall effect in Sec.~\ref{sec:boundary_qhe_tdflux},
we now discuss the possibility of regarding a heat transfer between two boundaries as the quantized thermal Hall current [Fig.~\ref{fig:laughlin_thermal}(b)]. 
When the both boundaries are in equilibrium at the same temperature $T$, the total free energy conserves due to (\ref{eq:freeenergy_leftright}), which indicates a heat is transferred between
boundaries by threading a gravitomagnetic flux.
The amount of the transferred heat is evaluated as
$
 \delta Q
 =
 T\delta S
 =
 -Td\delta F/d T
$.
By imposing the continuity equation of the heat
at the left boundary, a thermal current in the bulk is determined by $Lj^T_y-\delta \dot{Q}=0$.
Therefore
\begin{align}
 j^T_y
 =
 \frac{\delta \dot{Q}}{L}
 =
 -\frac{T}{L}
 \frac{d\delta\dot{F}}{dT}
 =
 \frac{\pi k_{\text{B}}^2T^2}{6\hbar}
 (-\dot{A}_x^\text{g}).
\end{align}
This expression indicates that, if we recognize the time derivative of the gravitomagnetic vector potential as a fictitious temperature gradient by $-T^{-1}\nabla_xT=-\dot{A}^\text{g}_x$, a heat transfer in the $y$ direction between two boundaries can also be regarded as a quantized thermal Hall current.
Notice that, in addition to the Tolman-Ehrenfest relation $-T^{-1}\nabla_xT=-\nabla_x \sigma$, a gravitational expression of a temperature gradient should be given by
\begin{align}
 -T^{-1}\nabla_xT=-\nabla_x \sigma - \dot{A}^\text{g}_x,
 \label{eq:fictitoiustemperaturegradient}
\end{align}
which is analogous to the expression of the electric field in terms of the electric potential $\phi$ and the vector potential $A$ in electromagnetism: $E_x=-\nabla_x \phi-\dot{A}_x$.
A similar expression has been employed in evaluation of the thermal current \cite{tatara15}, although definition of the vector potential in this literature is different from ours.

\section{Bulk argument for the quantized thermal Hall effect}
\label{sec:bulk}

In this final section,
we will develop yet another 
argument for the quantized thermal Hall effect
following the spirit of the original Laughlin's 
argument presented in 
Sec.\
\ref{Laughlin's original argument for the quantum Hall effect}. 
We will apply 
the modular transformation (\ref{eq:modular_transformation}) 
to the bulk electronic states forming the Landau levels, and examine an adiabatic transport induced by the modular transformation.
As in the original Laughlin's argument, 
our discussion here relies on and is limited to 
single-particle eigenfunctions of the Landau levels,
but 
gives a complementary view to the boundary argument presented in
Sec.\ \ref{sec:boundary_the}.

\subsection{Modular transformations for bulk wavefunctions}
\label{sec:bulk_wavefunction_modulartransformation}

Consider the Fourier modes of 
the fermion field on the Euclidean (2+1)-dimensional spacetime
labeled by the fermionic Matsubara frequency 
$\omega_n=2\pi(n+1/2)/\hbar\beta\,(n\in\mathbb{Z})$ 
and the momentum $k_x=2\pi l/L\,(l\in \mathbb{Z})$,
\begin{align}
 \tilde{\psi}(i\omega_n,k_x,y)
 =
 \frac{1}{\sqrt{\beta L}}
 \int d^2x
 e^{i\omega_n t-ik_x x}
 \psi(t,x,y).
\end{align}
Consider a continuous diffeomorphism of the base manifold as a function of the threaded gravitomagnetic flux.
Boundary conditions (\ref{eq:bc_before}) and (\ref{eq:bc_after}) are continuously connected by an intermediate boundary condition
\begin{align}
 \begin{aligned}
  &\psi(t+s\hbar\beta,x+L,y)=e^{-s(2m+1)\pi i}\psi(t,x,y),
  \\
  &\psi(t+\hbar\beta,x,y)=-\psi(t,x,y),
 \end{aligned}
 \label{eq:bc_intermediate}
\end{align}
where $m$ is an arbitrary integer and $s=\Phi^\text{E}/\Phi^\text{E}_0\in[0,1]$.
The fermion field satisfying (\ref{eq:bc_intermediate}) can be expanded by plain waves $\exp[-i\omega_n(t-s\hbar\beta x/L)+ik_x^{(s)}x]$ where $k_x^{(s)}=2\pi l/L-s(2m+1)\pi/L\,(l\in\mathbb{Z})$.
The modular transformation (\ref{eq:modular_transformation}) transforms a Fourier mode continuously as
\begin{align}
 &\tilde{\psi}(i\omega_n,k_x^{(0)},y) \notag\\
 &\to
 \frac{1}{\sqrt{\beta L}}
 \int d^2x
 \exp \left[
 i\omega_n
 \left(
 t
 -
 \frac{s\hbar\beta x}{L}
 \right)
 -
 ik_x^{(s)} x
 \right]
 \psi(t,x,y) \notag\\
 &=
 \frac{1}{\sqrt{\beta L}}
 \int d^2x
 \exp \left[
 i\omega_n
 t
 -
 i
 \left(
 k_x^{(s)}
 +
 \frac{s\omega_n\hbar\beta}{L}
 \right) 
 x
 \right]
 \psi(t,x,y) \notag\\
 &=
 \tilde{\psi}(i\omega_n,k_x^{(0)}+s(n-m)2\pi/L,y),
 \label{eq:fourier_gravitomagneticflux}
\end{align}
At $s=1$, the momentum is changed as $k_x\to k_x+(n-m)2\pi/L$.
Thus, by expanding 
the fermion field
with respect to the imaginary time, 
the modular transformation results in 
a frequency-dependent momentum shift.
One can remove an integer $m$ by threading magnetic flux quanta.
This prescription does not affect the following argument, since the magnetic flux does not induce the thermal Hall current.
For later convenience, we consider twice the unit of the modular transformation ($s=2$), and the momentum is shifted as $k_x\to k_x+(2n+1)2\pi/L$.
As explained in Sec.~\ref{Laughlin's original argument for the quantum Hall effect}, a momentum shift in the quantum Hall state is accompanied with an adiabatic 
shift of the center of mass of wavefunctions, 
which can be read off from 
(\ref{eq:position_momentum}) as
\begin{align}
 y\to
 y+(2n+1)\delta y,
 \label{eq:position_shift_grav}
\end{align}
where $\delta y=2\pi\hbar/|e|BL$.
Thus, by threading the gravitomagnetic flux $2\Phi^\text{E}_0$,
bulk quantum Hall electronic states with the Matsubara frequency $\omega_n$ are adiabatically transferred from 
their original localized positions to 
their $(2n+1)$th neighboring positions.
This should be contrasted with the original Laughlin's argument 
for the quantum Hall effect,
where, after threading a magnetic flux quantum $\Phi_0$, 
all electronic states are equally shifted to their neighboring positions.
When the quantum Hall system is in a thermal equilibrium, the gravitomagnetic-flux threading leaves the whole electronic system unchanged.

\subsection{The quantized thermal Hall current induced by 
the static gravitational potential}

Consider the situation that a temperature gradient is applied uniformly in the bulk.
Local temperature is defined through 
Luttinger's trick using the Tolman-Ehrenfest relation
$
 T(y)
 =
 \bar{T}
 e^{\sigma(y)}
$,
where $e^{-2\sigma}=g_{00}$, and $\bar{T}$ is a reference temperature independent of location, and simultaneously serves as a bulk temperature when $\sigma(y)$ is small enough.
Here we focus on a specific position $y_0$ of a localized position of the Landau level wave function determined by (\ref{eq:position_momentum}).
The $j$th neighboring localized position deviated from $y_0$ is denoted by $y_j=y_0+j\delta y$.
Also, we define the local temperature at a position $y_j$ by $T_j\equiv T(y_j)$, and its inverse by $\beta_j\equiv (k_{\text{B}}T_j)^{-1}$.

In order to capture qualitatively 
the changes in physical quantities induced by an adiabatic shift (\ref{eq:position_shift_grav}), we consider the partition function of the bulk quantum Hall states under a uniform temperature gradient.
We assume that the position-dependent temperature is represented in the partition function by the upper bound of the imaginary time integral as $\beta(y)=(k_{\text{B}}T(y))^{-1}$.
Then the action and the partition function are given by
\begin{align}
 S
 &=
 \int d\bm{x}
 \int_0^{\hbar\beta(y)}dt\,
 \bar{\psi}(t,\bm{x})
 \left(
 \hbar\partial_t
 +
 \mathcal{H}
 -
 \mu
 \right)
 \psi(t,\bm{x}),
 \nonumber \\
Z&= \int \mathcal{D}\bar{\psi} \mathcal{D}\psi\,
\exp (-S/\hbar),
 \label{eq:partitionfunction_temperaturegradient}
\end{align}
where $\bm{x}=(x,y)$, and $\mathcal{H}$ is the Hamiltonian of the bulk two-dimensional electron system under a perpendicular magnetic field, defined in (\ref{eq:2deg}).
The fermion field operator is expanded by the eigenstate wavefunctions of the Landau levels (\ref{eq:qh_wavefunction}) as  
\begin{align}
 &\psi(t,\bm{x})
 =
 \sum_{N,k_x}
 \phi_{N,k_x}(\bm{x})
 \frac{1}{\sqrt{\beta_j}}
 \sum_n
 e^{-i\omega_n(y_j)t}
 \psi_{n,N,k_x}, \\
 &\bar{\psi}(t,\bm{x})
 =
 \sum_{N,k_x}
 \phi_{N,k_x}^{\ast}(\bm{x})
 \frac{1}{\sqrt{\beta_j}}
 \sum_n
 e^{i\omega_n(y_j)t}
 \bar{\psi}_{n,N,k_x},
\end{align}
where $\omega_n(y_j)=(2n+1)\pi/\hbar\beta_j$, and $\beta_j$ is uniquely determined by $k_x$.
Then the action becomes
\begin{align}
 S/\hbar
 =
 \sum_{n,N,k_x}
 \bar{\psi}_{n,N,k_x}
 \left(
 -i\hbar \omega_n(y_j)
 +
 \epsilon_N
 -
 \mu
 \right)
 \psi_{n,N,k_x},
\end{align}
where $\epsilon_{N}$ is the $N$th Landau level energy (\ref{eq:landaulevel_energy}), and we have used the fact that the Landau level wavefunction $\phi_{N,k_x}$ is localized about the position $y_j$.
Thus we decompose the partition function by the momentum $k_x$ and calculate the path integral part by part as
\begin{align}
 Z 
 &=
 \prod_{n,N,j}
 \beta_j
 \left(
 -i\hbar\omega_n(y_j)
 +
 \epsilon_N
 -
 \mu
 \right),
 \label{eq:partitionfunction_temperaturegradient}
\end{align}
where the summation over the momentum $k_x$ is replaced by that over the index of the localized position $j$.
The total free energy is given by
\begin{align}
 F
 &=
 -\sum_{n,N,j}
 \beta_j^{-1}
 \ln
 \left[
 \beta_j\left(
 -i\hbar \omega_n(y_j)
 +
 \epsilon_N
 -
 \mu
 \right)
 \right]
 \equiv
 \sum_{n,j}
 F_n(y_j).
 \label{eq:partitionfunction_decomposition}
\end{align}

\begin{figure}
 \centering
 \includegraphics[width=52mm]{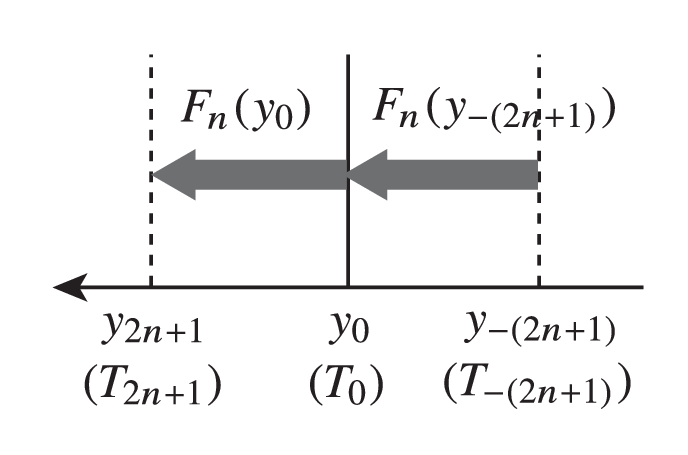}
 \caption{Transferred components of the bulk free energy in a quantum Hall state induced by the gravitomagnetic flux $\Phi^g=-2i\hbar\beta$. \label{fig:bulk_freeenergytransport}}
\end{figure}

Let us now focus on the local free energy at position $y_0$.
A local change of the bulk free energy can be evaluated by collecting parts of the partition function localized at $y_0$ before and after threading the gravitomagnetic flux.
Before threading the gravitomagnetic flux, the local free energy at $y_0$ is given by
\begin{align}
 F(y_0)
 &=
 \sum_n
 F_n(y_0).
\end{align}
Consider threading a uniform gravitomagnetic flux $\Phi^\text{E}$.
As we showed in Sec.~\ref{sec:bulk_wavefunction_modulartransformation}, when the flux $\Phi^\text{E}=2\hbar\beta_j$ is threaded, a Fourier mode with $(\omega_n(y_j),k_x)$, which is localized at $y_j$, is adiabatically changed to a mode with $(\omega_n(y_j),k_x+(2n+1)2\pi/L)$.
As for the local free energy at $y_0$,
a part of the free energy with $\omega_n(y_0)$
originally at $y_0$
flows out to $y_{2n+1}$ 
when $\Phi^\text{E}=2\hbar\beta_0$.
On the other hand, 
the free energy 
with $\omega_n(y_{-(2n+1)})$
at $y_{-(2n+1)}$
flows into $y_0$ when $\Phi^\text{E}=2\hbar\beta_{-(2n+1)}$ (Fig.~\ref{fig:bulk_freeenergytransport}).
Then the local free energy change at $y_0$ is given by 
\begin{align}
 \delta F(\Phi^\text{g}; y_0)
 &=
 i\Phi^\text{g}
 \sum_n
 \left[
 \frac{F_n(y_{-(2n+1)})}
 {2\hbar\beta_{-(2n+1)}}
 -
 \frac{F_n(y_{0})}{2\hbar\beta_0}
 \right],
 \label{eq:localpartitionfunction_change}
\end{align}
where we assume $\delta F$ to be a smooth function of the gravitomagnetic flux $\Phi^\text{g}$.

The right-hand side of (\ref{eq:localpartitionfunction_change}) is evaluated as follows.
Assuming the temperature gradient is relatively small, one obtains
\begin{align}
 &\delta F(\Phi^\text{g}; y_0) 
 \simeq
 i\Phi^\text{g}
 \sum_{n}
 \left(
 T_{-(2n+1)}-T_0
 \right)
 \left(
 \frac{\partial}{\partial T}
 \frac{F_n(y)}{2\hbar\beta}
 \right)_{y_0} \notag\\
 &
 =
 i\Phi^\text{g}
 \delta T
 \sum_{n}
 (2n+1)
 \frac{\partial}{\partial T}
 \sum_{N}
 \frac{
 \ln
 \left[
 \beta\left(
 -i\hbar\omega_n
 +
 \epsilon_N
 -
 \mu
 \right)
 \right]}
 {2\hbar\beta^2}
\notag\\
 &=
 -
 \frac{\Phi^\text{g} \delta T}
 {2\pi\hbar}
 \sum_N
 \frac{\partial}{\partial T}
 \beta^{-1}
 \sum_n
 (-i\hbar\omega_n)
 \ln
 \left[
 \beta
 \left(
 -i\hbar\omega_n
 +
 \epsilon_N
 -
 \mu
 \right)
 \right],
\end{align}
where $\delta T=T_{j+1}-T_j$ is the difference of the temperature between neighboring localized positions.
Evaluating the Matsubara summation, one obtains
\begin{align}
 \sum_n
 (-i\hbar\omega_n)
 \ln
 \left[
 \beta
 \left(
 -i\hbar\omega_n
 +
 \epsilon_N
 -
 \mu
 \right)
 \right]
 =
 G
 \left(
 \epsilon_N-\mu
 \right),
 \label{eq:matsubarasum}
\end{align}
where $G(z)$ is the integral of $\beta z/(e^{\beta z}+1)$.
At low temperatures, $G(z)$ is expanded with respect to the temperature by the Sommerfeld expansion as
\begin{align}
 G(z)
 =
 \int_{\infty}^z dz'
 \frac{\beta z'}{e^{\beta z'}+1}
 =
 \theta(-z)
 \left(
 \frac{\beta z^2}{2}
 -
 \frac{\pi^2}{6\beta}
 \right)
 +
 O(T^3),
\end{align}
where $\theta$ is the Heaviside step function.
The local free energy change is then given by
\begin{align}
 \delta F(\Phi^\text{g}; y_0)
 =
 \nu
 \frac{\pi k_{\text{B}}^2T_0}{6\hbar}
 \Phi^\text{g}\delta T,
\end{align}
where $\nu$ is the number of filled Landau levels and is equal to the total Chern number of filled energy levels.
Since each localized position is separated by an interval $\delta y$, the bulk thermal current is given by
\begin{align}
 j_x^T
 =
 \frac{1}{L\delta y}
 \frac{\partial \delta F(\Phi^\text{g}; y_0)}{\partial A_x^\text{g}}
 =
 \nu
 \frac{\pi k_{\text{B}}^2T_0}{6\hbar}
 \nabla_y T,
 \label{eq:the_bulk_result}
\end{align}
where $\delta T=(\nabla_y T) \delta y$ and $\Phi^\text{g}=LA_x^{\text{g}}$ are used.
The above relation is the quantized thermal Hall effect in the quantum Hall state with the Chern number $\nu$.
(\ref{eq:the_bulk_result}) satisfies the Wiedemann-Franz law with the Laughlin's original result (\ref{eq:laughlinoriginal}).
The above argument quantitatively describes how a thermal Hall current can flow adiabatically in a gapped bulk.

\section{Conclusion}
\label{sec:conclusion}

We studied the generalization of Laughlin's magnetic-flux-threading argument 
to the quantized thermal Hall effect in terms of gravity,
from the perspective of both bulk and boundary theories.

The boundary argument reveals that the global diffeomorphism anomaly
accounts for the quantized thermal Hall effect.
More precisely, we formulated, quantitatively,
the responses of the chiral boundary modes against the gravitomagnetic flux,
by making use of the global diffeomorphism anomaly.
The boundary modes gain or lose their free energy
during threading the gravitomagnetic flux
depending on the central charge and the temperature.
We have shown that this anomaly explains the quantized thermal Hall effect.
When boundaries are in contact with heat baths at different temperatures, 
the thermal Hall current flows in the direction perpendicular 
to the temperature difference,
and is quantized in units of the chiral central charge.

Guided by the very precise analogy between
the Laughlin's original argument for the charge transport and its thermal version,
which holds at the level of edge theories, 
we further discussed the corresponding bulk picture:
The Landau level states respond to the gravitomagnetic flux
by adiabatic shift of their localized positions,
the distance of which is dependent on the Matsubara frequency.
We evaluated the change in the free energy under
threading of the gravitomagnetic flux,
and further related it to the quantized thermal Hall
current carried by the bulk electronic states.
Although as we have shown there is an almost exact parallelism between the
thermal transport at the level of quantum anomalies, 
the precise nature of
the free energy generation by the {\it frequency-dependent} adiabatic motion of electrons in the Landau level
is still somewhat mysterious
(as compared to the charge pumping by the adiabatic motion of Landau orbits).
It is an important future problem to study the nature of the free energy
generation more precisely.

Finally, we again stress that our free energy is defined only globally due to the global nature of large diffeomorphism.
This should be contrasted with effective field theory descriptions
which are local (e.g., see Refs.\ \onlinecite{stone12,bradlyn15}).
As noted earlier\cite{stone12},
the gravitational Chern-Simons term is not able to describe the
response which could be generated by the finite gravito potential and gravitomagnetic potential.
In this paper (see also Refs.\ \onlinecite{nomura12,nakai16}), we attempted to derive the finite temperature
effective action different from the gravitational Chern-Simons theory.
Within the physics of edge theories, we have derived (1+1)-dimensional
the effective action describing the thermal transport edge theory.
The result is consistent with the
known result (``the replacement rule'')
in the context of the chiral magnetic effect (and the related
field).
The possible bulk effective field theory, consistent with the boundary 
effective theory, is presented in Ref.\ \onlinecite{nomura12}.

\acknowledgements
The authors acknowledge S. Fujimoto, and N. Yokoi for helpful discussions.
This work is supported by 
World Premier International Research Center
Initiative (WPI),
Grant-in-Aid for Scientific
Research (No.~JP15H05854 and No.~JP26400308) from the Ministry of Education, Culture, Sports, Science and Technology (MEXT), Japan,
and 
the National Science Foundation grant DMR-1455296.

\appendix

\section{Gravitomagnetic flux in metric and periodicity}
\label{sec:equivalence}

We consider the metric of the (2+1)-dimensional Euclidean spacetime under the gravitomagnetic vector potential.
Reduction of the following argument to the (1+1)-dimensional case is apparent.
The spacetime metric is given by
\begin{align}
 g_{\mu\nu}
 =
 \begin{pmatrix}
  1 & A^\text{E}_x & 0 \\
  A^\text{E}_x & 1+{A^\text{E}_x}^2 & 0 \\
  0 & 0 & 1
 \end{pmatrix},
\end{align}
and the corresponding frame field $\underline{\bm{e}}_{\mu}$ by
\begin{align}
 \underline{\bm{e}}_0
 =
 \begin{pmatrix}
  1 \\ 0 \\ 0
 \end{pmatrix}, \quad
 \underline{\bm{e}}_1
 =
 \begin{pmatrix}
  A^\text{E}_x \\ 1 \\ 0
 \end{pmatrix}, \quad
 \underline{\bm{e}}_2
 =
 \begin{pmatrix}
  0 \\ 0 \\ 1
 \end{pmatrix},
\end{align}
which satisfies $g_{\mu\nu}=\underline{\bm{e}}_{\mu}\cdot\underline{\bm{e}}_{\nu}$.
The coframe field $\bm{e}^{\mu}$, which is dual to the frame field, is given by
\begin{align}
 \bm{e}^0
 =
 \begin{pmatrix}
  1 \\ -A^\text{E}_x \\ 0
 \end{pmatrix}, \quad
 \bm{e}^1
 =
 \begin{pmatrix}
  0 \\ 1 \\ 0
 \end{pmatrix}, \quad
 \bm{e}^2
 =
 \begin{pmatrix}
  0 \\ 0 \\ 1
 \end{pmatrix},
\end{align}
which satisfies $\bm{e}^{\mu}\cdot\underline{\bm{e}}_{\nu}=\delta^{\mu}_{\nu}$, and $g_{\mu\nu}(\bm{e}^{\mu})_{\alpha}(\bm{e}^{\nu})_{\beta}=\delta_{\alpha\beta}$.

Here we show that one can cancel the gravitomagnetic vector potential in the metric by a diffeomorphism of the spacetime torus given by (\ref{eq:spacetime_transformation}), as long as the gravitomagnetic vector potential is uniform. 
The coframe field couples to the covariant derivative to make it invariant under the general coordinate transformation, as $(\bm{e}^{\mu})_{\alpha}D_{\mu}$.
Since the gravitomagnetic vector potential is constant in (imaginary) time and space, the spin connection $\omega_{\mu}$ vanishes.
The covariant derivative is rewritten as
\begin{align}
 (\bm{e}^{\mu})_{\alpha}
 \left(
 \hbar\partial_{\mu}
 -
 ieA_{\mu}
 \right)
 &=
 \hbar\partial'_{\alpha}
 -
 ieA'_{\alpha},
 \label{eq:cancel_metric}
\end{align}
where $\partial'_{\alpha}=(\bm{e}^{\mu})_{\alpha}\partial_{\mu}$, and $A'_{\alpha}=(\bm{e}^{\mu})_{\alpha}A_{\mu}$.
The new coordinate $x'$ resulting from the gravitomagnetic flux is given in terms of the original coordinate $x$ as
\begin{align}
 (t',x',y')
 &=
 (t
 +
 A^{\text{E}}_xx,x,y),
 \label{eq:app_coordinateTransformation}
\end{align}
which agrees with the diffeomorphism (\ref{eq:spacetime_transformation}).

When the quantum of the gravitomagnetic flux $\Phi^{\text{E}}_0=\hbar\beta$ is threaded, the transformation (\ref{eq:app_coordinateTransformation}) leads to the change of the boundary condition from
\begin{align}
 &\psi(x^0+\hbar\beta,x^1,x^2)
 =
 -\psi(t,x,y) \\
 &\psi(t,x+L,y)
 =
 \psi(t,x,y)
\end{align}
on the region $A$ defined by $t\in[0,\hbar\beta],x\in[0,L],y\in[-\infty,\infty]$ to
\begin{align}
 &\psi(t'+\hbar\beta,x',y')
 =
 -\psi(t',x',y')\\
 &\psi(t'+\hbar\beta,x'+L,y')
 =
 -\psi(t',x',y')
\end{align}
on the region $A'$ defined by $t'\in[\hbar\beta x'/L,\hbar\beta(1+x'/L)],x'\in[0,L],y'\in[-\infty,\infty]$.
Then solving the eigenvalue problem of the Lagrangian density with the gravitomagnetic flux
\begin{align}
 \hat{L}[A^\text{g}_x]
 \psi_a(x)
 =
 (i\omega_n-\epsilon_a)
 \psi_a(x)
\end{align}
on the undistorted region $A$ is equivalent to the same problem without the gravitomagnetic flux
\begin{align}
 \hat{L}[A^\text{g}_x=0]
 \psi_a(x')
 =
 (i\omega_n-\epsilon_a)
 \psi_a(x')
 \label{eq:lagrangian_eigenvalue_wog}
\end{align}
on the distorted region $A'$.

The Lagrangian density operator of the Dirac fermion under the electromagnetic vector potential and the gravitomagnetic vector potential is
\begin{align}
 \hat{L}[A^{\text{g}},A]
 &=
 \sqrt{g}
 \left[
 i\hbar v_{\text{F}}
 \underline{\gamma}^{\mu}
 D_{\mu}
 -
 m
 \right],
 \label{eq:lagrangian_dirac_emg}
\end{align}
where the gamma matrix on the curved spacetime $\underline{\gamma}^{\mu}$ satisfies $\{\underline{\gamma}^{\mu},\underline{\gamma}^{\nu}\}=2g^{\mu\nu}$, which is related to the one on the flat spacetime $\gamma^{\alpha}$ via $\underline{\gamma}^{\mu}= (\bm{e}^{\mu})_{\alpha}\gamma^{\alpha}$, where $\{\gamma^{\alpha},\gamma^{\beta}\}=2\delta^{\alpha\beta}$.
The identity (\ref{eq:cancel_metric}) transforms the derivative in (\ref{eq:lagrangian_dirac_emg}) as
\begin{align}
 \underline{\gamma}^{\mu}
 \left(
 \hbar\partial_{\mu}
 -
 ieA_{\mu}
 \right)
 &=
 \gamma^{\alpha}
 \left(
 \hbar\partial'_{\alpha}
 -
 ieA'_{\alpha}
 \right).
 \label{eq:dirac_cancel}
\end{align}
Due to $\sqrt{g}=(\text{det}[g_{\mu\nu}])^{1/2}=1$,
(\ref{eq:dirac_cancel}) cancels the gravitomagnetic vector potential, and the remaining problem is to solve the equation of the form (\ref{eq:lagrangian_eigenvalue_wog}).

In a similar way, the quadratic Hamiltonian under the uniform gravitomagnetic vector potential\cite{bradlyn15}
\begin{align}
 \hat{L}[A^{\text{g}},A]
 &=
 \sqrt{g}
 \left[
 \frac{i}{2}(\bm{e}^{\mu})_{0}
 D_{\mu}
 +
 \frac{1}{2m}
 (\bm{e}^{\mu})_{a}
 (\bm{e}^{\nu})^{a}
 D_{\mu}
 D_{\nu}
 \right]
\end{align}
can also be transformed to the problem (\ref{eq:lagrangian_eigenvalue_wog}).

\end{document}